%% file: main.tex
\documentclass[12pt]{article}

\usepackage{float}

\usepackage[date=year,style=numeric,giveninits=true,sorting=none]{biblatex}
\AtEveryBibitem{\clearlist{language} \clearlist{location} }
\renewbibmacro{in:}{}

\usepackage[dvipdfm,margin=1.1in]{geometry}
\usepackage{amsfonts,mathtools,amsthm,bm}
\usepackage{subfiles}
\usepackage{subcaption}
\usepackage[dvipdfmx]{graphicx}
\usepackage{authblk}

\newtheorem{theorem}{Theorem}[section]
\newtheorem{definition}[theorem]{Definition}
\newtheorem{corollary}[theorem]{Corollary}
\newtheorem{lemma}[theorem]{Lemma}

\newcommand{\inpr}[2]{\left\langle #1\, ,\, #2\right\rangle}
\newtheorem{model}{Model}

\title{Strongly trapped space-inhomogeneous \\
quantum walks in one dimension}
\author[1]{Chusei Kiumi}
\author[2]{Kei Saito}
\affil[1]{\footnotesize Mathematical Science Unit, Graduate School of Engineering Science, Yokohama National University, Hodogaya, Yokohama, 240-8501, Japan}
\affil[2]{Department of Information Systems Creation, Faculty of Engineering, Kanagawa University, Kanagawa, Yokohama, 221-8686, Japan}
\date{\empty}

\begin{document}
\maketitle
\vspace{-1.2cm}
\subfile{abstract}

\subfile{introduction}
\subfile{definition}
\subfile{model}

\begin{figure}[H]
\centering
\begin{subfigure}[H]{0.49\textwidth}
\centering
\includegraphics[width=0.92\linewidth]{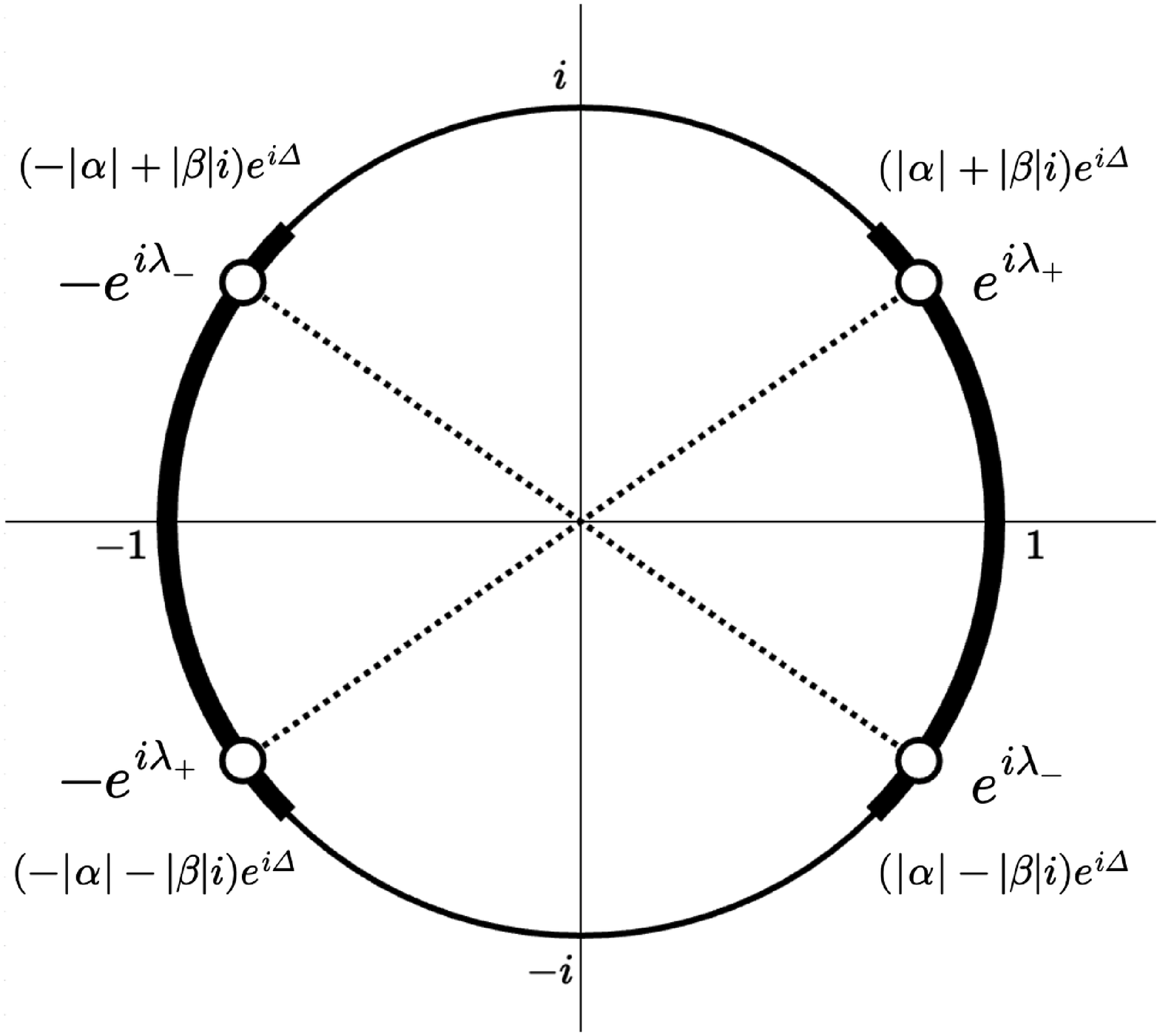} 
\caption{}
\end{subfigure}
\begin{subfigure}[H]{0.49\textwidth}
\centering
\includegraphics[width=0.95\linewidth]{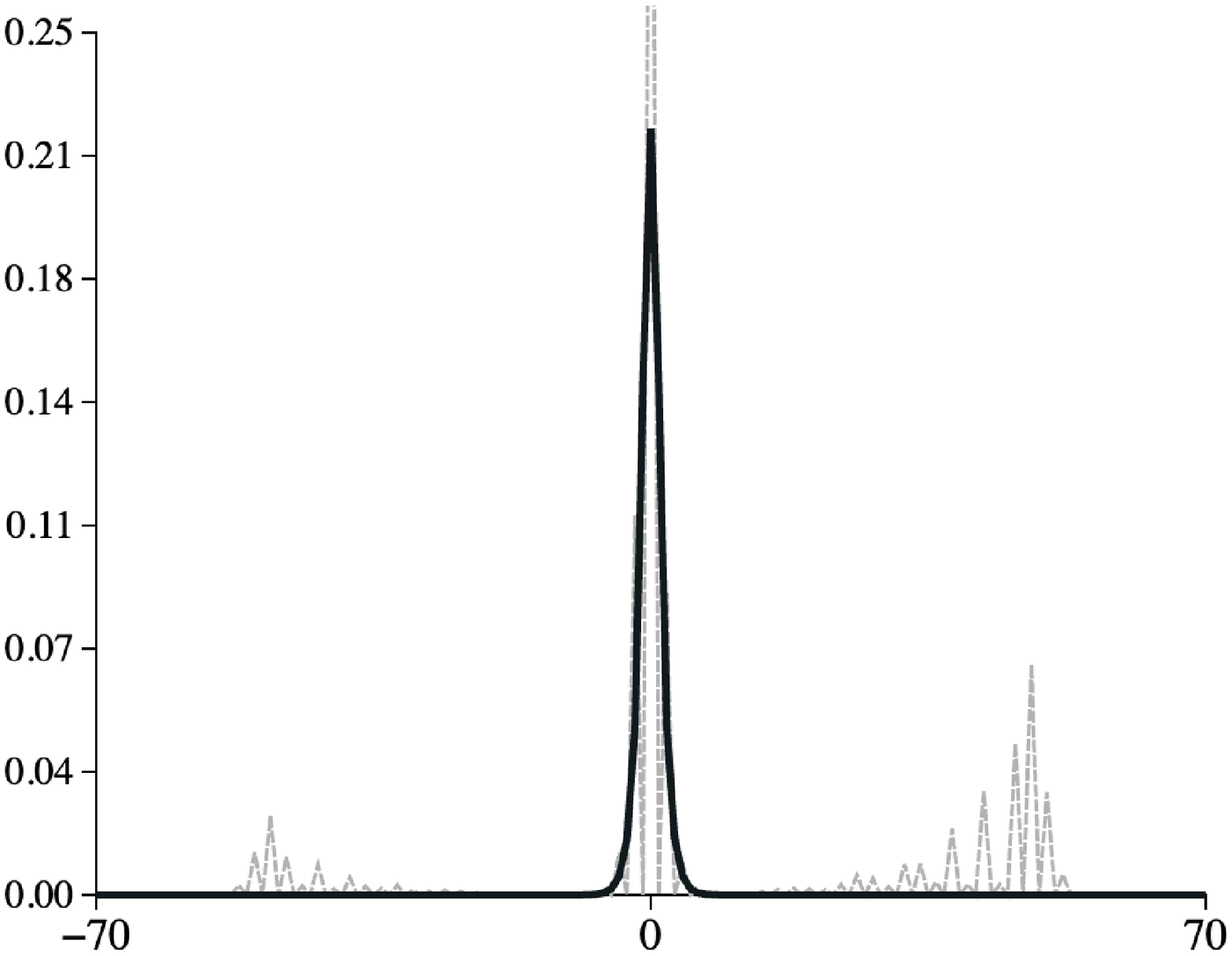}
\caption{}
\end{subfigure}
\caption{
Example of Model \ref{THEO_MODEL1} with parameters $\Delta _{o} =\Delta =0, \alpha =\alpha _{o} =\beta  =\frac{1}{\sqrt{2}}, \beta_{o}=\frac{i}{\sqrt{2}}, \psi _{1} =\frac{1}{\sqrt{2}}, \psi _{2} =\frac{1}{\sqrt{2}}$, which is strongly trapped. (a) illustrates the eigenvalues, and the bold lines indicate the range of the existence of eigenvalues when $\vert  \beta\vert  $ is fixed. (b) shows the probability distribution at time $t=70$ (dotted line) and the time-averaged limit distribution (bold black line), where the horizontal axis indicates position $x$.}
\label{fig:1}  
\end{figure}

\begin{figure}[H]
\begin{subfigure}[H]{0.49\textwidth}
\centering
\includegraphics[width=0.92\linewidth]{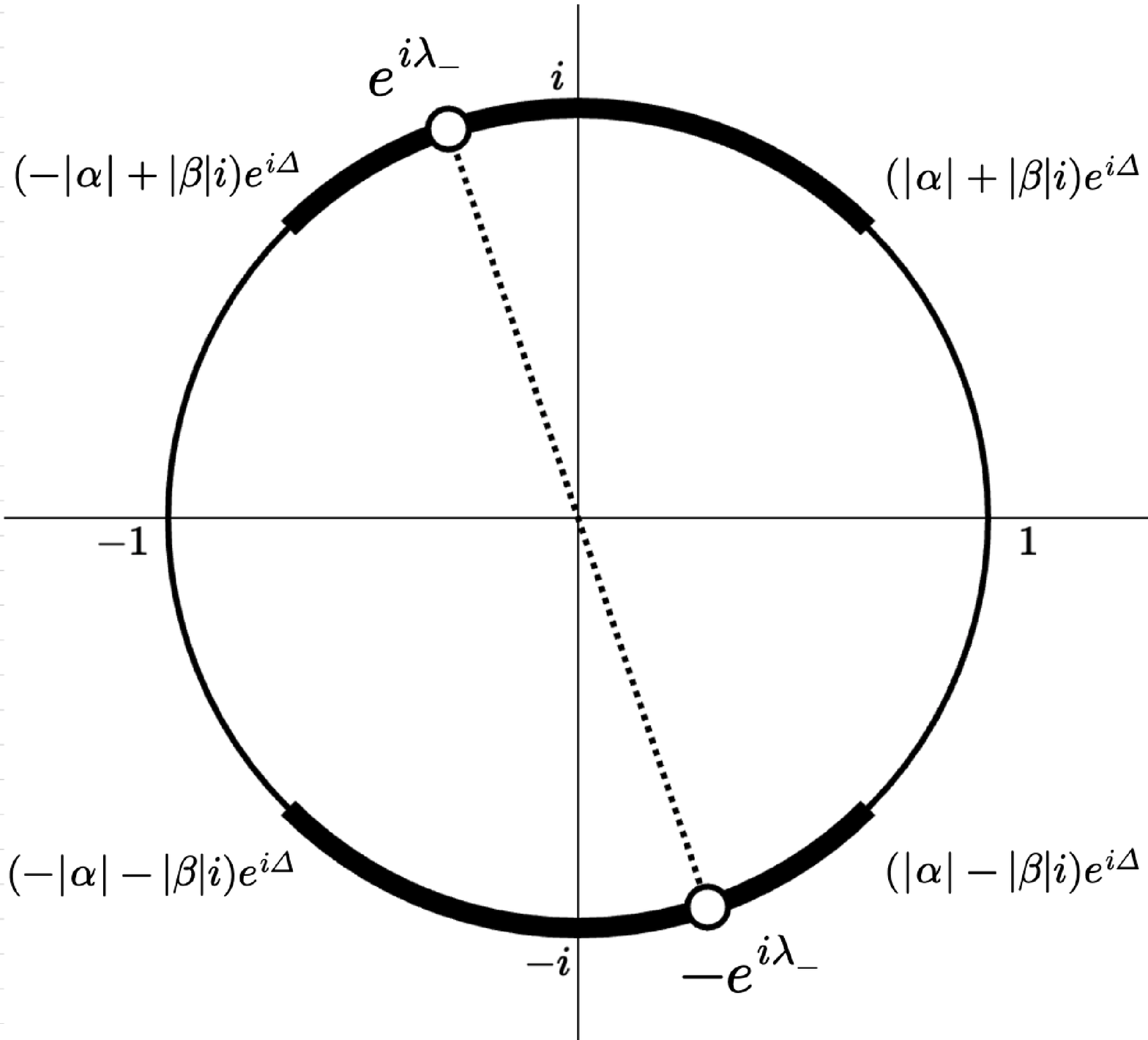} 
\caption{}
\end{subfigure}
\begin{subfigure}[H]{0.49\textwidth}
\centering
\includegraphics[width=0.95\linewidth]{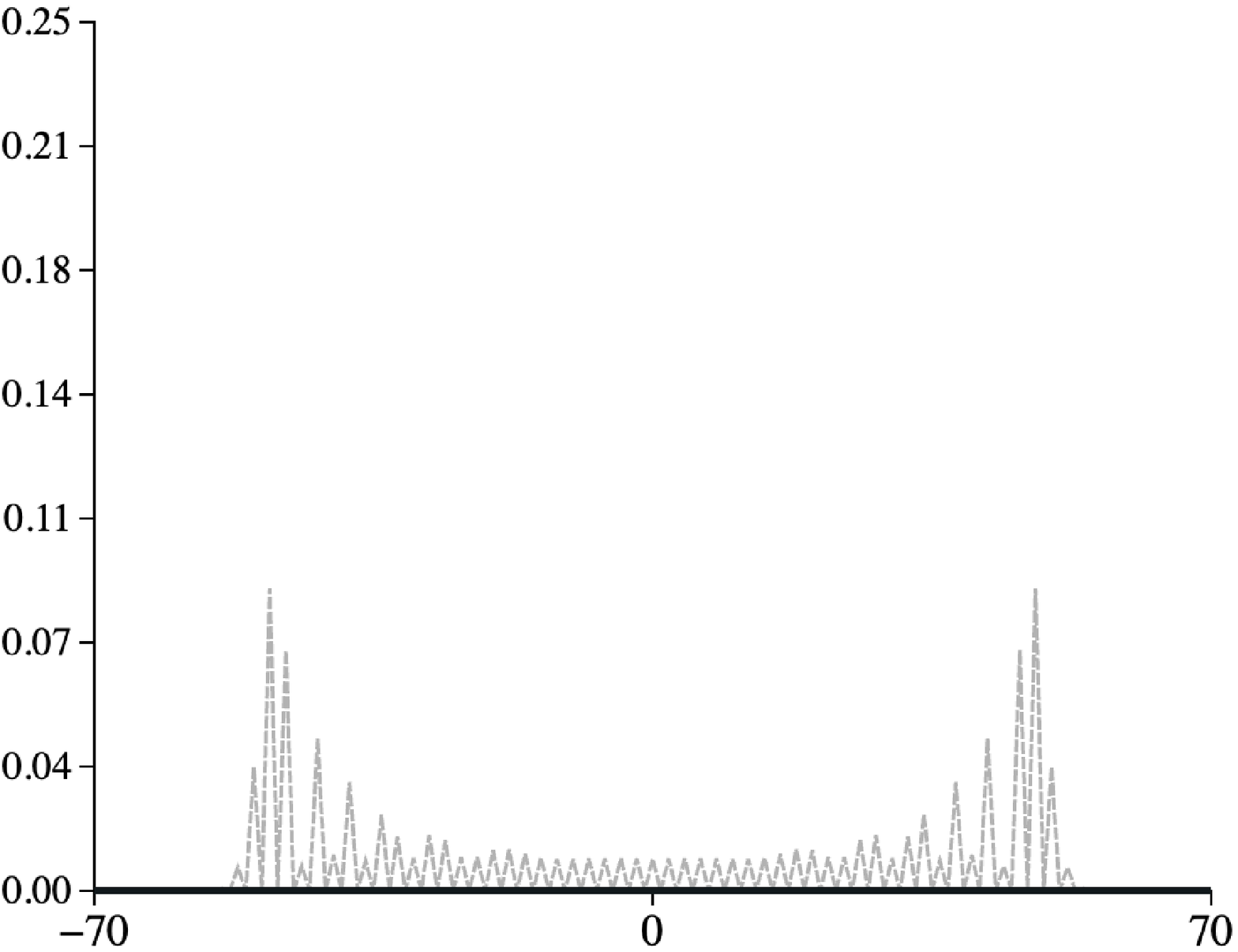}
\caption{}
\end{subfigure}
\caption{Example of Model \ref{THEO_MODEL2} with parameters $\Delta _{o} =0,\Delta =\frac{\pi }{2} ,\ \alpha =\beta =\alpha _{o} =\beta _{o} =\frac{1}{\sqrt{2}} ,\psi _{1} =\frac{1}{\sqrt{2}} ,\psi _{2} =\frac{-i}{\sqrt{2}}$, which is {\bf not} strongly trapped. (a) illustrates the eigenvalues, and the bold lines indicate the range of the existence of eigenvalues when $\vert  \beta\vert  $ is fixed. (b) shows the probability distribution at time $t=70$ (dotted line), where the horizontal axis indicates position $x$. In this case, the time-averaged distribution becomes 0 for any $x\in\mathbb{Z}$ with this specific initial state.}
\label{fig:2}  
\end{figure}

\begin{figure}[H]
\begin{subfigure}[H]{0.49\textwidth}
\centering
\includegraphics[width=0.92\linewidth]{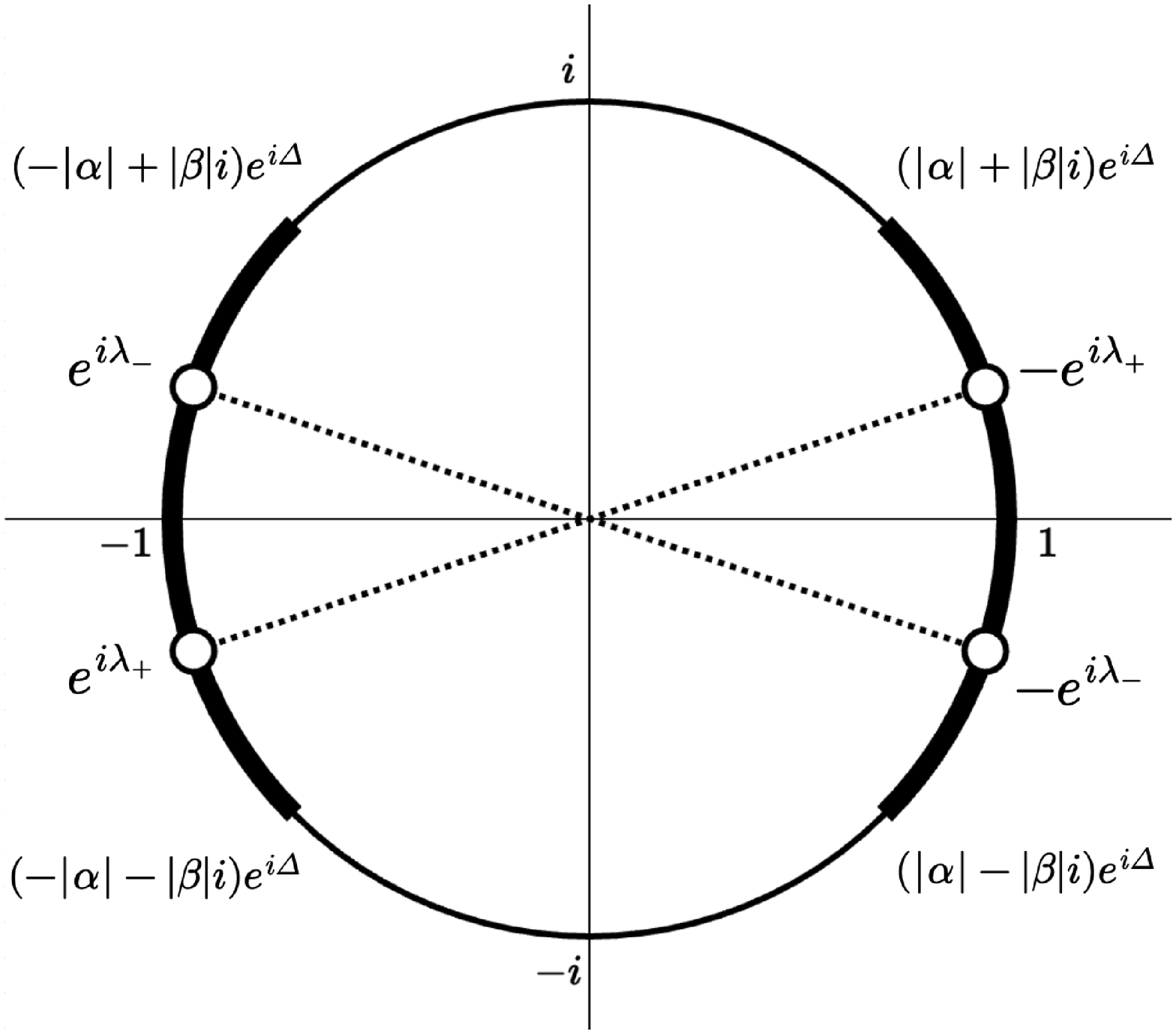} 
\caption{}
\end{subfigure}
\begin{subfigure}[H]{0.49\textwidth}
\centering
\includegraphics[width=0.95\linewidth]{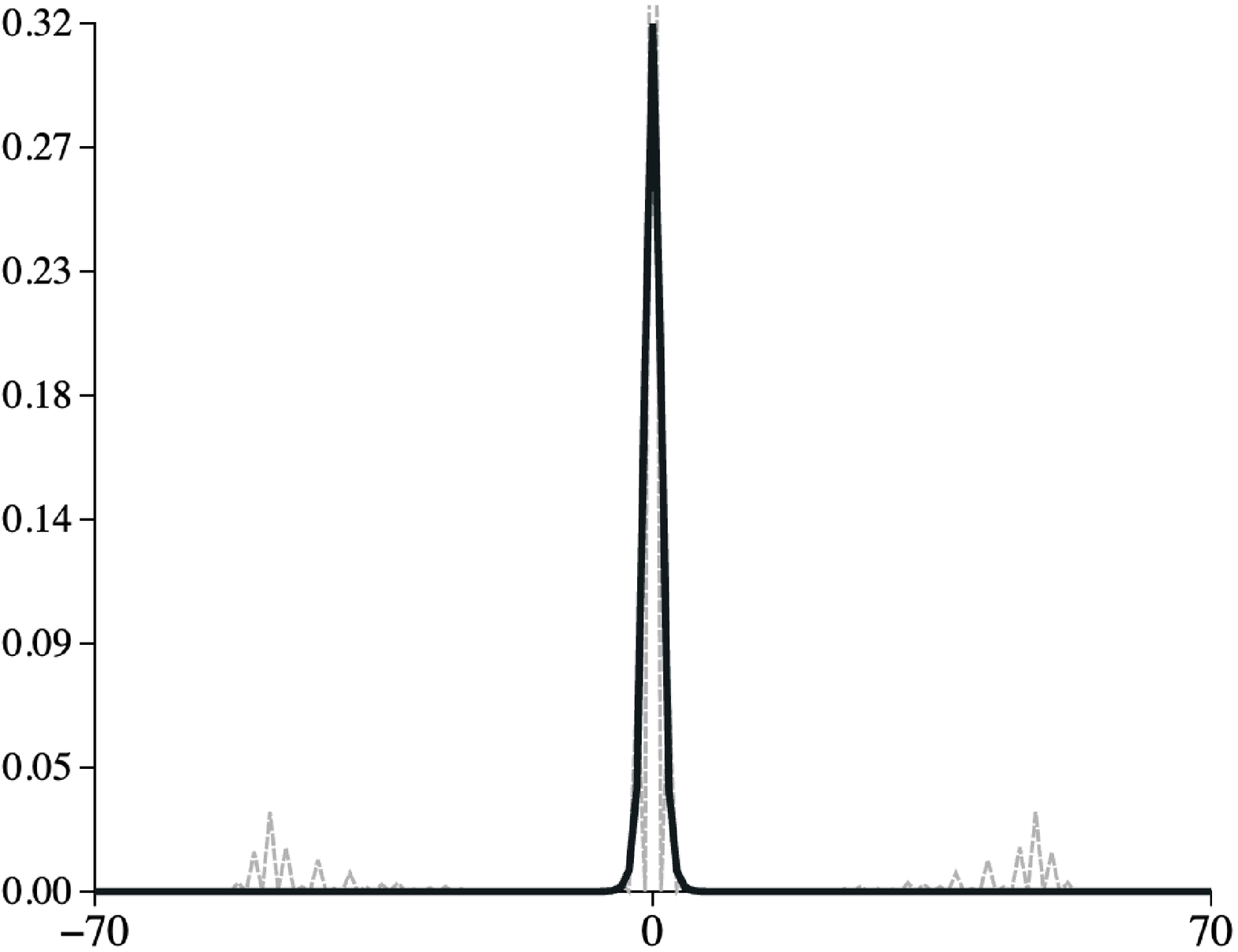}
\caption{}
\end{subfigure}
\caption{Example of Model \ref{THEO_MODEL2} with parameters $\Delta _{o} =0,\Delta  =\pi ,\ \alpha =\beta =\alpha _{o} =\beta _{o} =\frac{1}{\sqrt{2}} ,\psi _{1} =\frac{1}{\sqrt{2}} ,\psi _{2} =\frac{-i}{\sqrt{2}}$ , which is strongly trapped. (a) illustrates the eigenvalues, and the bold lines indicate the range of the existence of eigenvalues when $\vert  \beta\vert  $ is fixed. (b) shows the probability distribution at time $t=70$ (dotted line) and the time-averaged limit distribution (bold black line), where the horizontal axis indicates position $x$.}
\label{fig:3}  
\end{figure}

\begin{figure}[H]
\begin{subfigure}[H]{0.49\textwidth}
\centering
\includegraphics[width=0.92\linewidth]{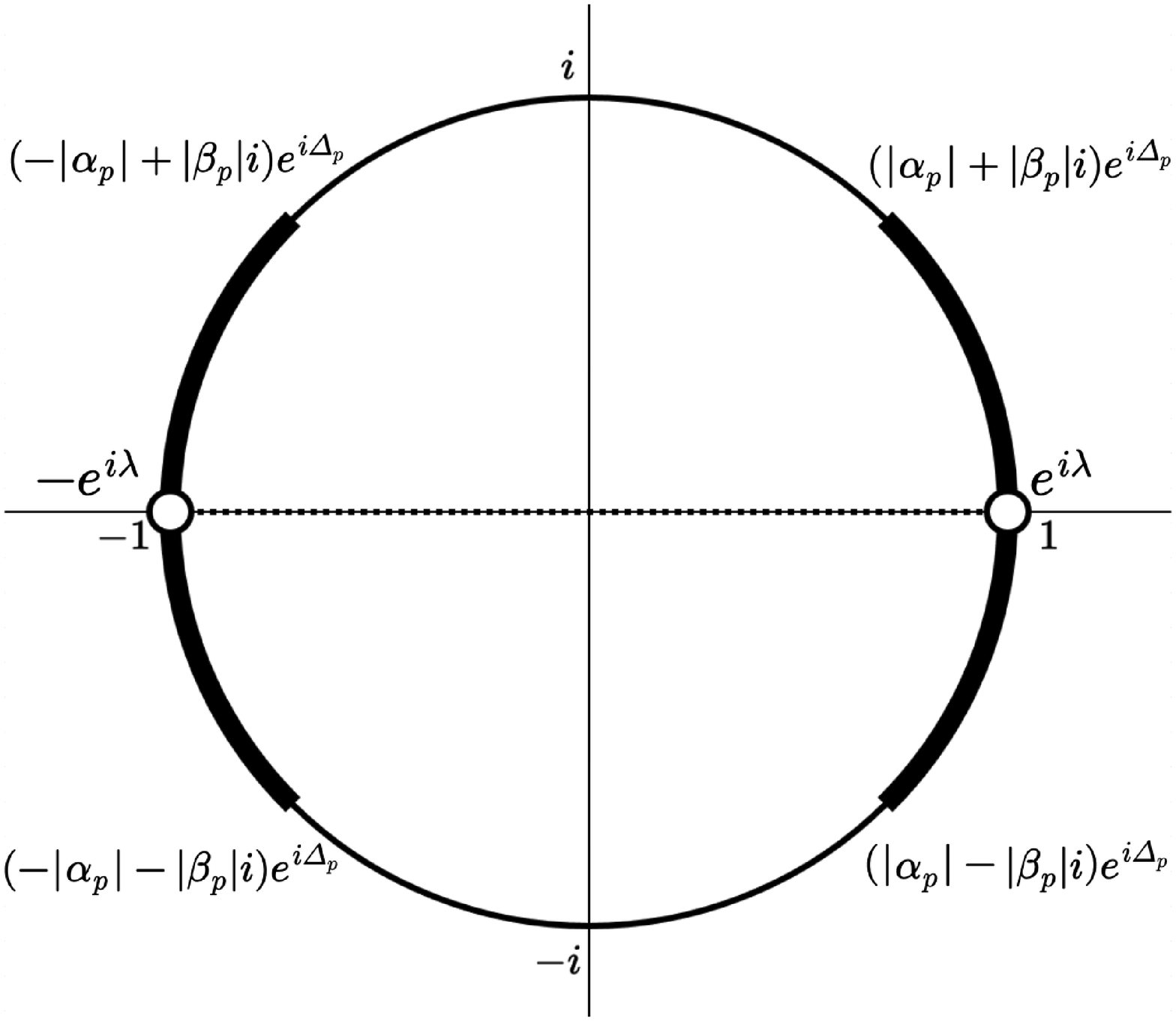} 
\caption{}
\end{subfigure}
\begin{subfigure}[H]{0.49\textwidth}
\centering
\includegraphics[width=0.95\linewidth]{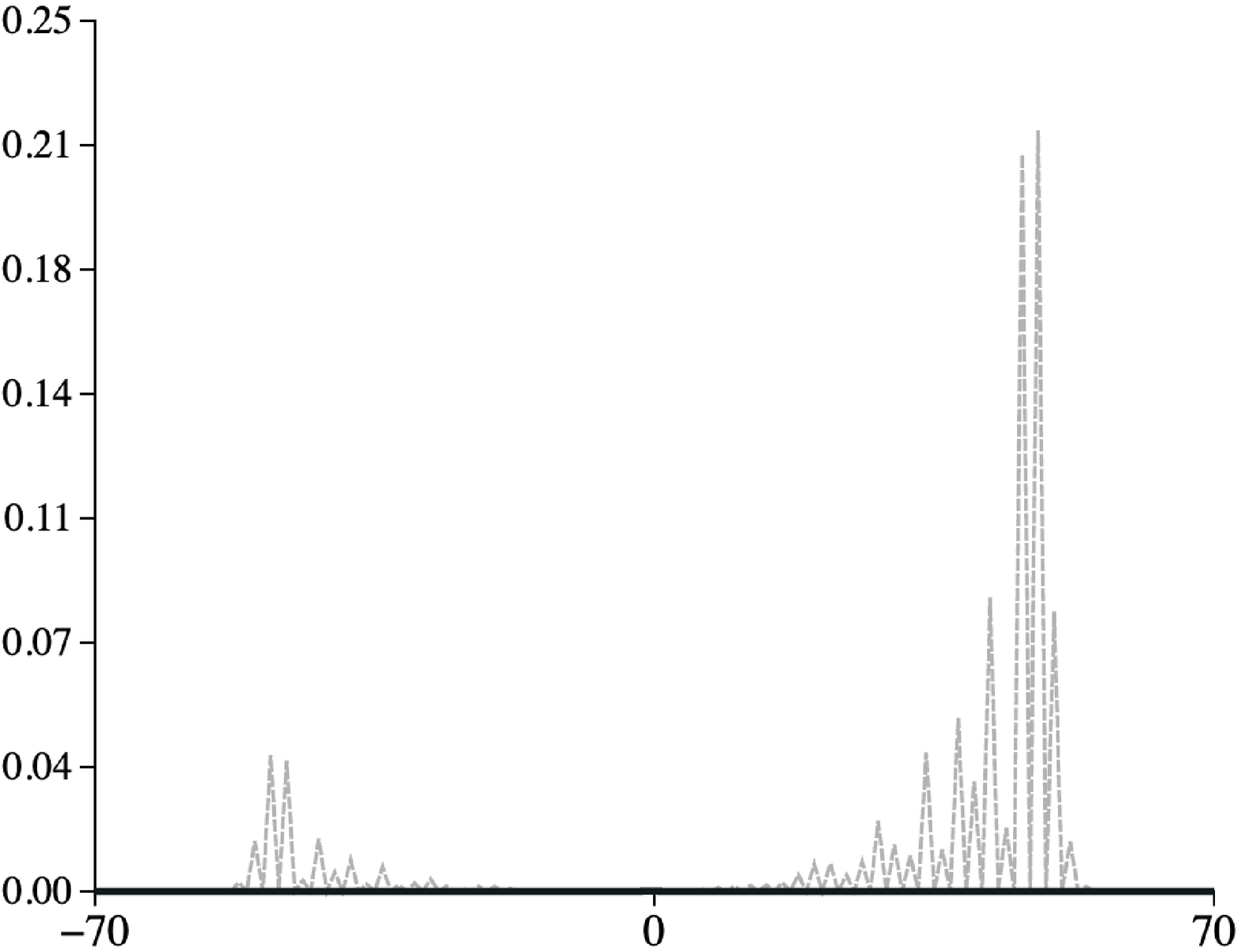}
\caption{}
\end{subfigure}
\caption{Example of Model \ref{THEO_MODEL3} with parameters $\Delta _{p} =\pi ,\Delta _{m} =0,\ \alpha _{p} =\beta _{p} =\alpha _{m} =\beta _{m} =\frac{1}{\sqrt{2}},\psi _{1} =\frac{1}{2}\sqrt{2+\sqrt{2}} ,\psi _{2} =-\frac{1}{2}\sqrt{2-\sqrt{2}}$, which is {\bf not} strongly trapped. (a) illustrates the eigenvalues, and the bold lines indicate the range of the existence of eigenvalues when $\vert  \beta_p\vert  $ is fixed. (b) shows the probability distribution at time $t=70$ (dotted line), where the horizontal axis indicates position $x$. In this case, the time-averaged distribution becomes 0 for any $x\in\mathbb{Z}$ with this specific initial state.}
\label{fig:4}  
\end{figure}

	\begin{figure}[H]
\begin{subfigure}[H]{0.49\textwidth}
\centering
\includegraphics[width=0.92\linewidth]{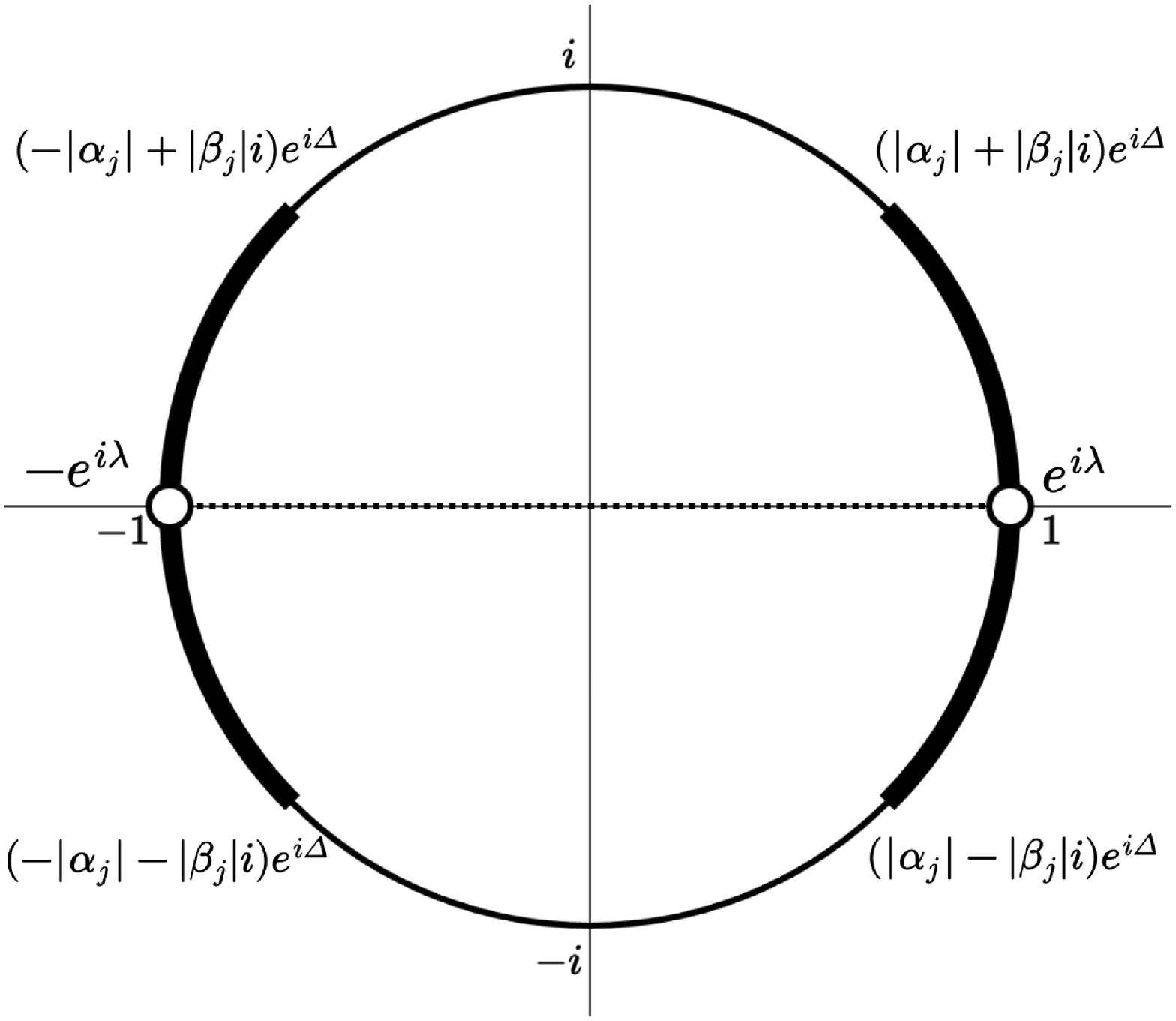} 
\caption{}
\end{subfigure}
\begin{subfigure}[H]{0.49\textwidth}
\centering
\includegraphics[width=0.95\linewidth]{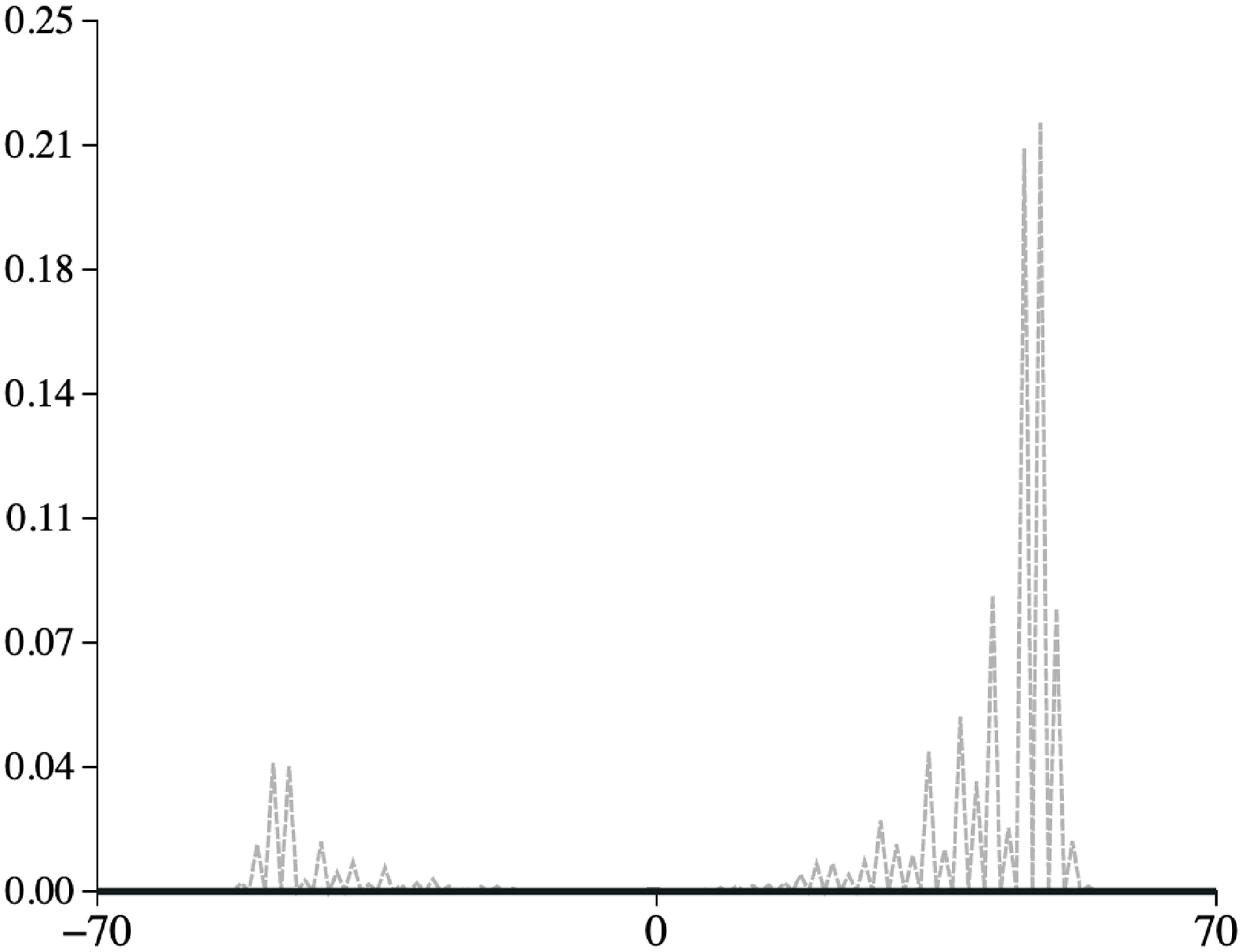}
\caption{}
\end{subfigure}
\caption{Example of Model \ref{THEO_MODEL4} with parameters $\Delta =0,\ \alpha _{p} =\alpha _{m} =\beta _{m} =\frac{1}{\sqrt{2}} ,\beta _{p} =-\frac{1}{\sqrt{2}} ,\ \phi _{1} =\frac{1}{2}\sqrt{2+\sqrt{2}} ,\phi _{2} =\frac{1}{2}\sqrt{2-\sqrt{2}}$, which is {\bf not} strongly trapped. (a) illustrates the eigenvalues, and the bold lines indicate the range of the existence of eigenvalues when $\vert  \beta_j\vert  \  (j=p\left(\left\lvert  \beta_{p}\right\rvert   \leq\left\lvert  \beta_{m}\right\rvert  \right),=m\left(\left\lvert  \beta_{m}\right\rvert  <\left\lvert  \beta_{p}\right\rvert  \right))$ is fixed. (b) shows the probability distribution at time $t=70$ (dotted line), where the horizontal axis indicates position $x$. In this case, the time-averaged distribution becomes 0 for any $x\in\mathbb{Z}$ with this specific initial state.}
\label{fig:5}  
\end{figure}

    \begin{figure}[H]
\begin{subfigure}[H]{0.49\textwidth}
\centering
\includegraphics[width=0.92\linewidth]{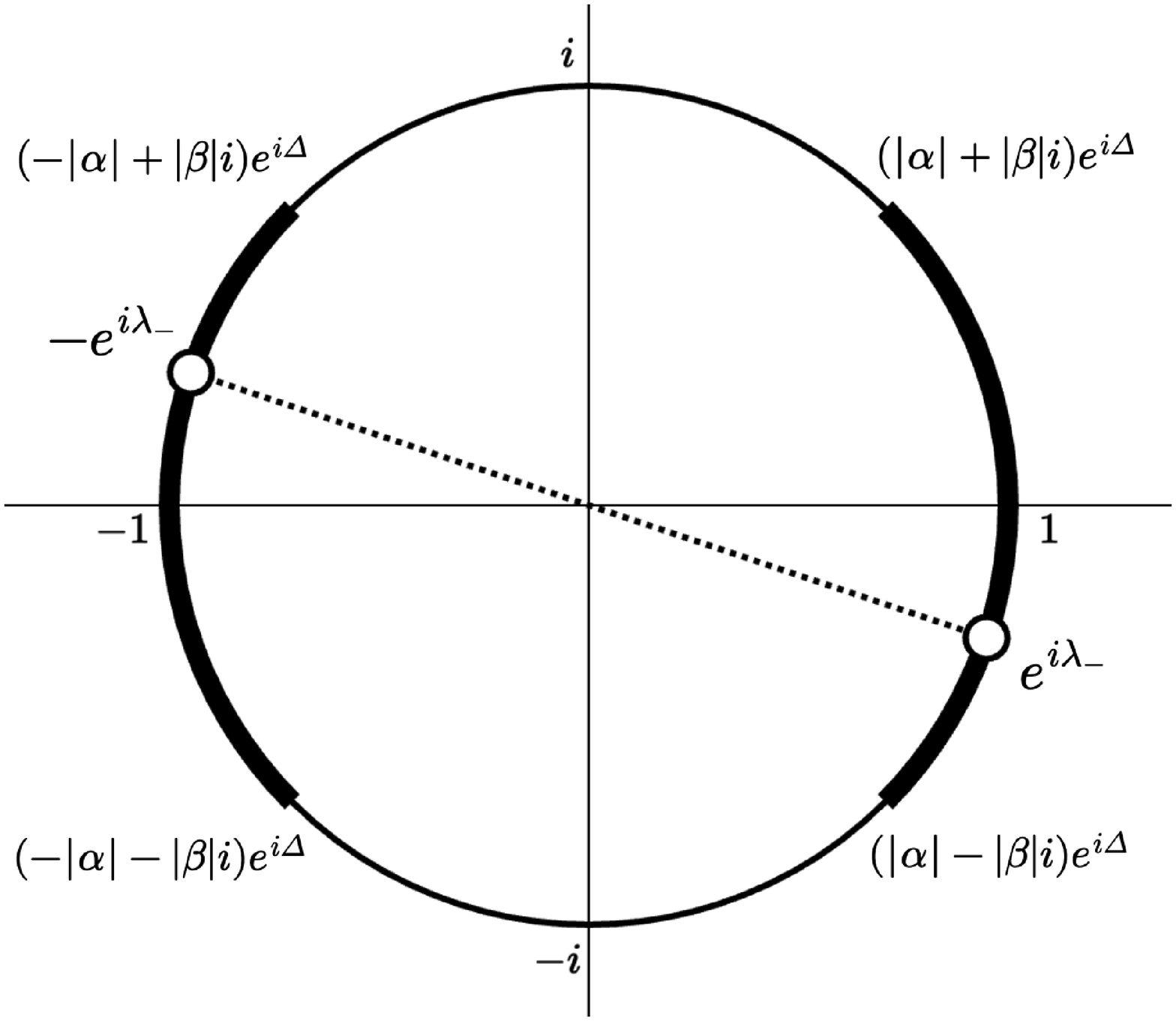} 
\caption{}
\end{subfigure}
\begin{subfigure}[H]{0.49\textwidth}
\centering
\includegraphics[width=0.95\linewidth]{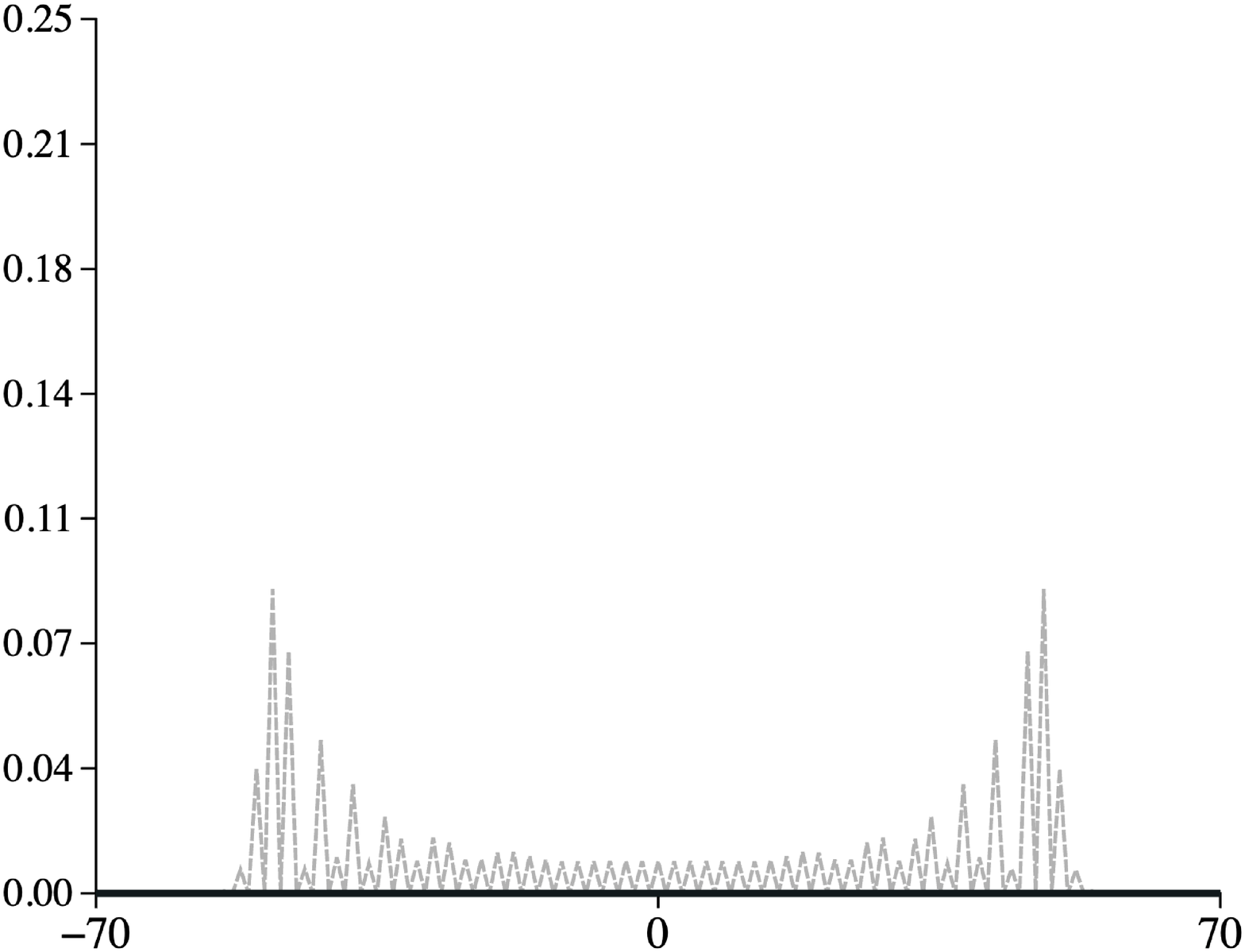}
\caption{}
\end{subfigure}
\caption{Example of Model \ref{THEO_MODEL5} with parameters $\Delta =\Delta _{o} =0,\alpha _{o} =1,\ \alpha _{p} =\alpha _{m} =\beta _{m} =\frac{1}{\sqrt{2}} ,\beta _{p} =\frac{i}{\sqrt{2}}, \psi _{1} =\frac{1}{\sqrt{2}} ,\psi _{2} =\frac{1}{\sqrt{2}} e^{i\frac{\pi }{4}}$, which is {\bf not} strongly trapped. (a) illustrates the eigenvalues, and the bold lines indicate the range of the existence of eigenvalues when $\vert  \beta\vert  $ is fixed. (b) shows the probability distribution at time $t=70$ (dotted line), where the horizontal axis indicates position $x$. In this case, the time-averaged distribution becomes 0 for any $x\in\mathbb{Z}$ with this specific initial state.}
\label{fig:6}  
\end{figure}
\begin{figure}[H]
\begin{subfigure}[H]{0.49\textwidth}
\centering
\includegraphics[width=0.92\linewidth]{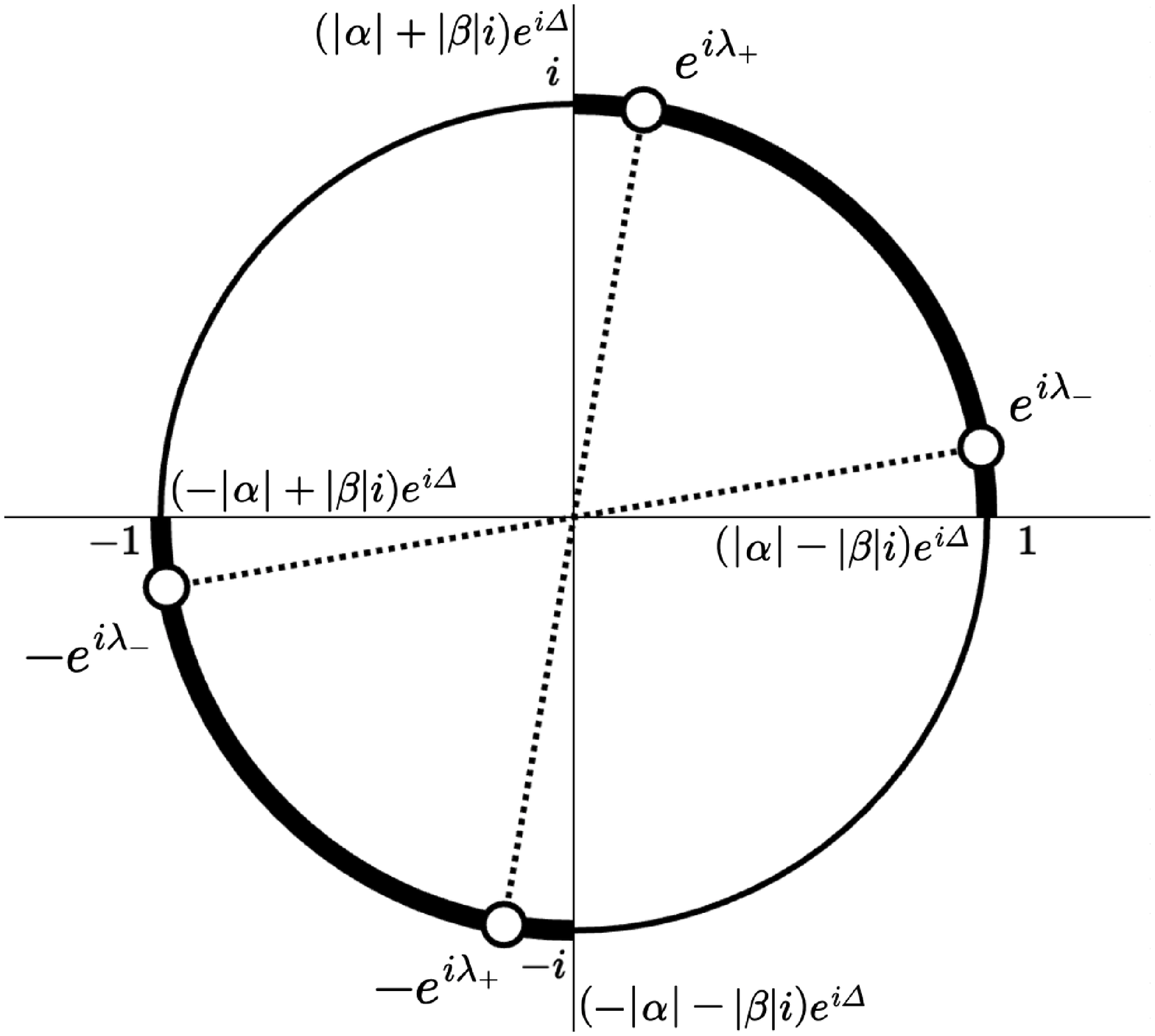} 
\caption{}
\end{subfigure}
\begin{subfigure}[H]{0.49\textwidth}
\centering
\includegraphics[width=0.95\linewidth]{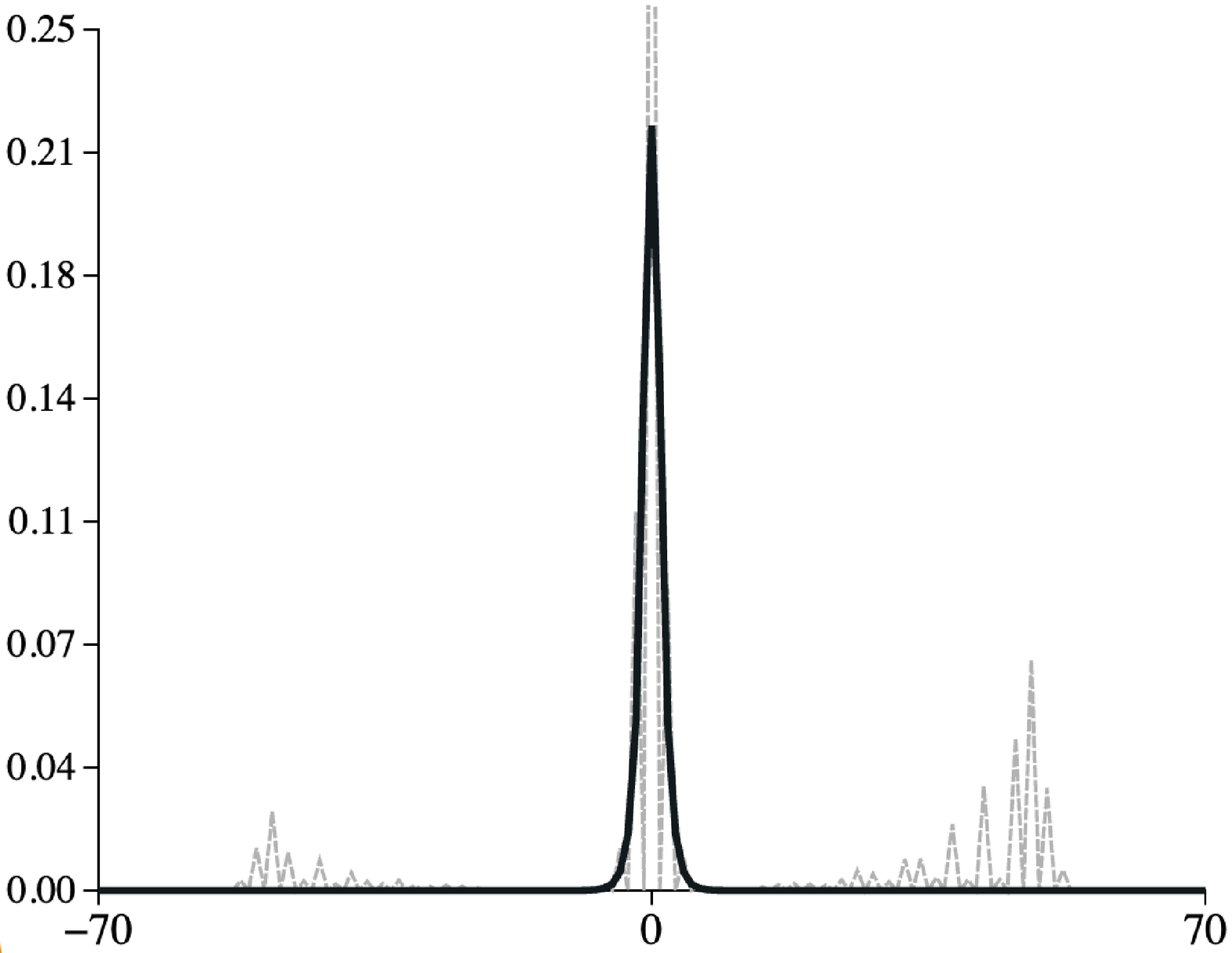}
\caption{}
\end{subfigure}
\caption{Example of Model \ref{THEO_MODEL5} with parameters $\Delta =\frac{\pi }{4} ,\Delta _{o} =0,\alpha _{o} =1,\ \alpha _{p} =\alpha _{m} =\beta _{m} =\frac{1}{\sqrt{2}} ,\beta _{p} =\frac{i}{\sqrt{2}},\psi _{1} =\frac{1}{\sqrt{2}} ,\psi _{2} =\frac{1}{\sqrt{2}}$ , which is strongly trapped. (a) illustrates the eigenvalues, and the bold lines indicate the range of the existence of eigenvalues when $\vert  \beta\vert  $ is fixed. (b) shows probability distribution at time $t=70$ (dotted line) and the time-averaged limit distribution (bold black line), where the horizontal axis indicates position $x$.}

\label{fig:7}  
\end{figure}

\subfile{summary}
\printbibliography

\section*{Appendix}
In the appendix, we give an abstract form of time-averaged limit distribution of two-phase quantum walks with one defect.
As denoted in Section 2.2, we remark that the time-averaged limit distribution with initial state $\Psi_0\in\mathcal{H}$ is written by the following.
\[
    \overline{\nu}_{\infty } (x) =
\sum _{e^{i\lambda} \in \sigma _{p} (U)}
\left|\inpr{\Psi^{\lambda }}{\Psi_0}\right|^2\ 
\|\Psi^{\lambda }(x)\|_{\mathbb{C}^2}^2,
\]
where $\Psi^\lambda \in \ker(U-e^{i\lambda})$.
For eigenvalue $e^{i\lambda}$, we put
\[
    M = 
    \frac{
    \alpha_m\zeta_m^> - e^{i(\lambda-\Delta_m)}
    }
    {\beta}
    e^{-i(\lambda - \Delta_o)}
\]
and normalized constant
\[
    N = 
    \left(
    \frac
    {(1-|\beta_o|^2)(1-|\zeta_p^<|^2)+(1+|\beta_o|^2)(1-|\zeta_m^<|^2)}
    {\left(1-|\zeta_p^<|^2\right)\left(1-|\zeta_m^<|^2\right)}
    +
    \frac
    {2\Re(\beta_oM)}
    {1-|\zeta_p^<|^2}
    \right)^{-1}
    .
\]
Then, associated eigenvector is
\begin{align*}
    \Psi^\lambda(x)
    =
    \sqrt{\dfrac{N}{2}}
    \begin{cases}
    \begin{bmatrix}
    e^{i(\lambda - \Delta_o)}(1+\beta_o M)
    \\[+5pt]
    -(\overline{\beta_o}+M)(\zeta_p^<)^{-1}
    \end{bmatrix}
    (\zeta_p^<)^x, \quad & x\geq 1,
    \\[+16pt]
    \begin{bmatrix}
    e^{i(\lambda - \Delta_o)}(1 + \beta_o M)
    \\[+5pt]
    -e^{i(\lambda -\Delta_o)}\alpha_oM
    \end{bmatrix}, \quad & x=0,
    \\[+16pt]
    \begin{bmatrix}
    \alpha_o\zeta_m^>
    \\
    -e^{i(\lambda-\Delta_o)}\alpha_oM
    \end{bmatrix}
    (\zeta_m^>)^x, \quad & x\leq -1.
    \end{cases}
\end{align*}
The key parts of the time-averaged limit distribution are
\begin{align*}
\|\Psi^\lambda(x)\|^2_{\mathbb{C}^2}
=
N^2
\begin{cases}
\left(
\dfrac{1+|\beta_o|^2}{2}+\Re(\beta_o M)
\right)
\left( 1+\left|\zeta_p^>\right|^2\right)
\left|\zeta_p^>\right|^{2x},\quad & x\geq 1,
\\[+5pt]
\left(1-\Re(\beta_oM)\right),\quad & x=0,
\\[+5pt]
\left(
\dfrac{1-|\beta_o|^2}{2}
\right)
\left(
1+\left|\zeta_m^>\right|^2
\right)
\left|\zeta_m^>\right|^{2x},\quad & x\leq -1,
\end{cases}
\end{align*}
and
\begin{align*}
    \left|
    \inpr{\Psi^\lambda}{\Psi_0}
    \right|^2
    =
    \frac{N^2}{2}
    {\bigg [}
    1 + \left(1+2\Re(\beta_o M)\right)|\psi_1|^2
    - 2\Re\left(
    (M + \overline{\beta_o})\alpha_o\overline{\psi_1}\psi_2
    \right)
    {\bigg ]},
\end{align*}
where $\Psi_0(x) =
    \begin{bmatrix}
    \psi_1 & \psi_2
    \end{bmatrix}^{t}$ with $|\psi_1|^2 + |\psi_2|^2 = 1$ if $x=0$, and $=
    \begin{bmatrix}
    0 & 0
    \end{bmatrix}^{t}$ if $x\neq 0$.
    Here, $t$ means the transpose operator.

\end{document}

%% file: abstract.tex
\begin{abstract}
  Localization is a characteristic phenomenon of space-inhomogeneous quantum walks in one dimension, where particles remain localized around their initial position. The existence of eigenvalues of time evolution operators is a necessary and sufficient condition for the occurrence of localization, and their associated eigenvectors are deeply related to the amount of localization, i.e., the probability that the walker stays around the starting position in the long-time limit. In a previous study by authors, the eigenvalues of two-phase quantum walks with one defect were studied using a transfer matrix, which focused on the occurrence of localization (Quantum Inf. Process 20(5), 2021). In this paper, we introduce the analytical method to calculate eigenvectors using the transfer matrix and also extend our results to characterize eigenvalues not only for two-phase quantum walks with one defect but also for a more general space-inhomogeneous model. 
 With these results, we quantitatively evaluate localization and study the strong trapping property by deriving the time-averaged limit distributions of five models studied previously.
\end{abstract}

%% file: introduction.tex
\section{Introduction}
Discrete-time quantum walks (QWs), which are quantum counterparts of random walks play important roles in the various fields. In this paper, we focus on two-state space-inhomogeneous QWs on the integer lattice \cite{Ambainis2001-os,Cantero2012-yk,Konno2005-uq}, where the dynamics of the particle depend on each position. Some classes exhibit an interesting property called localization. In these classes, the probability of finding
the particle around the initial location remains positive in the long-time limit. In particular, when this property occurs for any initial states starting from the origin, we say that the model is strongly trapped. The term ``strong trapping'' is introduced and the possible application for the trapping of light is mentioned in \cite{Kollar2015-ar}.  Moreover, the condition of strong trapping of two-dimensional homogeneous QWs is studied in \cite{Kollar2020-jv}. Mathematical analysis of localization is important for the manipulation of particles. For instance, localization on one-defect QWs where the walker at the origin behaves differently is used for quantum searching algorithms \cite{Ambainis2005-ha,Childs2004-xa,Shenvi2003-jw}, and localization on two-phase QWs where the walker behaves differently in each of negative and non-negative parts is related to the research of topological insulators \cite{Kitagawa2010-su,Endo2015-db}. Thus, in this paper, we primarily focus on localization and strong trapping on two-phase QWs with one defect including both one-defect and two-phase models. 

It is known that the occurrence of localization is equivalent to the existence of eigenvalues of the time evolution operator, and the amount of localization is deeply related to eigenvectors \cite{Segawa2016}. In the previous research by authors \cite{Kiumi2021-yg}, the necessary and sufficient condition of the occurrence of localization was given, and authors derived concrete eigenvalues for five models of two-phase QWs with one defect. Four of them are extensions of models in the previous studies \cite{Endo2020-or,Endo2015-db,Endo2014-bu,Endo2015-cy,Wojcik2012-kr}. However, how much the quantum walker localized around the initial position in the long-time limit depends on the initial states, and it is not clarified. Therefore, we are interested in the asymptotic behaviors and the class of strong trapping for two-phase QWs with one defect. In this paper, we reintroduce the method for the eigenvalue problem using the transfer matrix introduced in \cite{Kiumi2021-yg} in a more general and simpler way. Then, we reveal the time-averaged limit distribution and the class of strongly trapped QWs by this method. The transfer matrix is also used for several previous studies \cite{Danaci2019-qw,Kawai2018-ry,Kawai2017-fn}.

The rest of this paper is organized as follows. In Section \ref{sec:2}, we introduce notations and definitions of our models. The eigenvalue analysis using a transfer matrix is also assigned in this section.
Subsequently, we introduce the mathematical definition of strong trapping with the time-averaged limit distribution. We also define and show further analysis of eigenvectors for two-phase QWs with one defect in the last subsection. In Section \ref{sec:3}, We show the results of the concrete calculation of the time-averaged limit distribution and clarify the class of strong trapping for five models of two-phase QWs with one defect studied by authors previously \cite{Kiumi2021-yg}.

%% file: definition.tex
\section{Definitions}
\label{sec:2}
\subsection{Quantum walks on the integer lattice}
We introduce our QW model on the integer lattice $\mathbb{Z}$.
Let $\mathcal{H}$ be a Hilbert space of the states as follows:
\[
\mathcal{H}=\ell^2(\mathbb{Z} ; \mathbb{C}^2) =
\left\{
\Psi : \mathbb{Z} \to \mathbb{C}^2\ \middle\vert\ \sum_{x\in\mathbb{Z}}\|\Psi(x)\|_{\mathbb{C}^2}^2 < \infty
\right\}.
\]
For $\Psi\in\mathcal{H}$, we write $\Psi(x)=\begin{bmatrix}\Psi_L(x) & \Psi_R(x)\end{bmatrix}^t\in\mathbb{C}^2$, where $t$ is the transpose operator.  The time evolution $U=SC$ is defined by the product of two unitary operators $S$ and $C$ on $\mathcal{H}$ defined as below:
 \begin{align*}
	(S\Psi)(x)=\begin{bmatrix}
	\Psi_L(x+1)	\\ \Psi_R(x-1)
	\end{bmatrix},
	\qquad
	(C\Psi)(x)
	=
	   C_x\Psi(x),
	\end{align*}where $\{C_x\}_{x\in\mathbb{Z}}$ is a sequence of $2\times2$ unitary matrices called coin matrices. We write $C_x$ as follow:
    \begin{align*}
	C_x = e^{i\Delta_x}
	\begin{bmatrix}
	\alpha_x & \beta_x
	\\
	-\overline{\beta_x} & \overline{\alpha_x}
	\end{bmatrix},
	\end{align*}where $\alpha_x, \beta_x\in\mathbb{C},\ \alpha_x\neq 0,\ \Delta_x\in [0,2\pi)$ and $\vert\alpha_x\vert^2+\vert\beta_x\vert^2=1$.
	Let $x_+,x_-\in \mathbb{Z}$ with $x_+>0,\ x_-<0$ and $C_\infty,\ C_{-\infty}$ be $2\times 2$ unitary matrices. 
	We treat the model whose coin matrices satisfy the following.
\[
C_x =
\begin{cases}
C_\infty ,\quad & x\in [ x_{+} ,\infty ),
\\
C_{-\infty} ,\quad & x\in ( -\infty ,x_{-}].
\end{cases}
\]
With initial state $\Psi_0\in\mathcal{H}\ (\|\Psi_0\|_{\mathcal{H}}^2=1)$, the probability distribution at time $t\in\mathbb{Z}_{\geq 0}$ is defined by  $\mu_t^{(\Psi_0)}(x)=\|(U^t\Psi_0)(x)\|_{\mathbb{C}^2}^2$, where $\mathbb{Z}_{\geq 0}$ is the set of non-negative integers. Here, we say that the QW exhibits localization if there exists an initial state $\Psi_0\in\mathcal{H}$ and a position $x_0\in\mathbb{Z}$ which satisfy $\limsup_{t\to\infty}\mu^{(\Psi_0)}_t(x_0)>0$. It is known that the QW exhibits localization if and only if there exists an eigenvalue of $U$, that is, there exists $\lambda\in[0, 2\pi)$ and $\Psi\in\mathcal{H}$  such that
\begin{align*}
	U\Psi=e^{i\lambda}\Psi.
	\end{align*}
$\sigma_p(U)$ denotes the set of $e^{i\lambda}$ satisfying this condition. Let $J$ be a unitary operator on $\mathcal{H}$ defined as	\begin{align*}
	    (J\Psi)(x)=
	    \begin{bmatrix}
	    \Psi_L(x-1) \\ \Psi_R(x)
	    \end{bmatrix},
	    \quad \Psi\in\mathcal{H},\ x\in\mathbb{Z}.
	\end{align*}
	The inverse of $J$ is given as
\begin{align*}
	   	(J^{-1}\Psi)(x)=
	    \begin{bmatrix}
	    \Psi_L(x+1) \\ \Psi_R(x)
	    \end{bmatrix},
	    \quad \Psi\in\mathcal{H},\ x\in\mathbb{Z}.
	\end{align*}
	Moreover, we introduce the transfer matrix $T_x(\lambda)$ for $\lambda\in[0,2\pi)$ and $x\in\mathbb{Z}$ as below:
	\begin{align*}
	T_x(\lambda)=
	\frac{1}{\alpha_x}
	\begin{bmatrix}
	e^{i(\lambda-\Delta_x)} & -\beta_x
	\\
	-\overline{\beta_x} & e^{-i(\lambda-\Delta_x)}
	\end{bmatrix}.
	\end{align*}
	Unless otherwise noted, we abbreviate the transfer matrix $T_x(\lambda)$ as $T_x$.
	The transfer matrix is a normal matrix, and its inverse matrix is
	\begin{align*}
	T_x^{-1}=
	\frac{\alpha_x}{\vert\alpha_x\vert^2}
	\begin{bmatrix}
	e^{-i(\lambda-\Delta_x)} & \beta_x
	\\
	\overline{\beta_x} & e^{i(\lambda-\Delta_x)}
	\end{bmatrix}.
	\end{align*}
	Furthermore, the transfer matrix plays an important role in describing the eigenvectors.
	That is, a map $\Psi : \mathbb{Z}\to\mathbb{C}^2$ satisfies $(U-e^{i\lambda})\Psi = 0$ if and only if $\Psi$ satisfies the following equation for $x\in\mathbb{Z}$ :
	\[
	    (J\Psi)(x+1) = T_x(J\Psi)(x).
	\]
	Thus, we can analyze the eigenvalues and eigenvectors by finding suitable $\varphi = \Psi(0)\in\mathbb{C}^2$ such that $\Psi\in\mathcal{H}=\ell^2(\mathbb{Z};\mathbb{C}^2)$.
	For the details, see our previous study \cite{Kiumi2021-yg}.
	
	\begin{corollary}
	\label{CORO_A}
	For $\lambda\in[0,2\pi)$ and $\varphi\in\mathbb{C}^2$, we define $\tilde\Psi : \mathbb{Z}\to \mathbb{C}^2$ as follows:
	\begin{align*}
\tilde{\Psi } (x) & =\left\{\begin{array}{ l l }
T_{x-1} T_{x-2} \cdots T_{1} T_{0} \varphi , & x >0,\\
\varphi , & x=0,\\
T^{-1}_{x} T^{-1}_{x+1} \cdots T^{-1}_{-2} T^{-1}_{-1} \varphi , & x< 0.
\end{array}\right. \\[+8pt]
 & =\begin{cases}
T_{\infty}^{x-x_{+}} T_{+} \varphi,  & x_{+} \leq x,\\
T_{x-1} \cdots T_{0} \varphi,  & 0< x< x_{+},\\
\varphi,  & x=0,\\
T^{-1}_{x} \cdots T^{-1}_{-1} \varphi,  & x_{-} < x< 0,\\
T^{x-x_{-}}_{-\infty } T_{-} \varphi,  & x \leq x_{-}.
\end{cases}
\end{align*}
where $T_{+}=T_{x_{+} -1} \cdots T_{0},\  T_{-} = T^{-1}_{x_{-}} \cdots T^{-1}_{-1}$ and $T_{\pm \infty} = T_{x\pm}$. Then, 
$\tilde \Psi$ is a unique solution of $(U-e^{i\lambda}) x = 0$ up to a constant multiple.
Moreover, the existence of non-trivial $\varphi$ satisfying $\tilde\Psi\in\mathcal{H}$ is a necessary and sufficient condition for $\ker(U-e^{i\lambda})\neq \{\mathbf{0}\}$, that is, $\Psi = J^{-1}\tilde \Psi\in\ker(U-e^{i\lambda})$.
    \end{corollary}
    
Let $\zeta_{x}^\pm$ be eigenvalues of $T_x$, then
    \begin{align*}
	\zeta_{x}^\pm=\frac{\cos(\lambda-\Delta_x)\pm\sqrt{\cos^2(\lambda -\Delta_x)-\vert\alpha_x\vert^2}}{\alpha_x}.
	\end{align*}
It is known that if $\cos^2(\lambda -\Delta_x)-\vert\alpha_x\vert^2\leq 0$, i.e., $\vert\zeta_{x}^\pm \vert = 1$, then $e^{i\lambda} \not \in \sigma_p(U)$. 
The detailed proof of this fact is in \cite{Kiumi2021-yg,Maeda2021-nk}.
Hence, we have the following lemma.
\begin{lemma}
\label{Lemma Lambda}
$\displaystyle \sigma _{p}( U) \subset \Lambda $ holds, where
\[\Lambda =\left\{e^{i\lambda} \in \mathbb{C}\ \middle\vert\ \vert\zeta ^{+}_{\pm \infty } \vert\neq \vert\zeta ^{-}_{\pm \infty }\vert,\ 
\zeta ^{+}_{\pm\infty},\zeta ^{-}_{\pm\infty}  \in \sigma _{p}( T_{\pm\infty})
\right\}.\]
\end{lemma}

For $e^{i\lambda}\in\Lambda$, this lemma guarantees that one of the absolute values of the eigenvalues of $T_x$ is strictly greater than $1$ and the other is strictly less than $1$ since $\vert\zeta^+_x\vert \vert\zeta^-_x\vert = 1$.
Here, without loss of generality, we can set $\zeta^{<}_\infty \in \sigma_p(T_\infty)$ (resp. $\zeta^{>}_{-\infty} \in \sigma_p(T_{-\infty})$), which satisfies $\vert\zeta^{<}_\infty\vert<1$ (resp. $\vert\zeta^{>}_{-\infty}\vert>1$).

\begin{theorem}
\label{Theorem Ker}
$e^{i\lambda} \in \sigma _{p}( U) $ if and only if the following two statements hold:
\begin{align*}
    &(1) e^{i\lambda} \in \Lambda.\\
    &(2) \ker\left(\left( T_{\infty} -\zeta ^{<}_{\infty }\right) T_{+}\right) \cap \ker\left(\left( T_{-\infty} -\zeta ^{>}_{-\infty }\right) T_{-}\right)\neq \{\mathbf{0}\}.
\end{align*}
Furthermore, $\Psi \in\ker(U-e^{i\lambda})$ if and only if $\Psi = J^{-1}\tilde \Psi$, where $\tilde \Psi$ is defined in Corollary \ref{CORO_A} with $\varphi \in \ker\left(\left( T_{\infty} -\zeta ^{<}_{\infty }\right) T_{+}\right) \cap \ker\left(\left( T_{-\infty} -\zeta ^{>}_{-\infty }\right) T_{-}\right)$.
\end{theorem}

\begin{proof}
If $e^{i\lambda}\in\sigma_p(U)$, Lemma \ref{Lemma Lambda} shows $e^{i\lambda} \in \Lambda$.
The eigenvector $\Psi$ associated with $e^{i\lambda}$ is expressed by $\varphi\in\mathbb{C}^2\setminus\{\mathbf{0}\}$ and transfer matrices from Corollary \ref{CORO_A}.
Then, since $\tilde \Psi = J \Psi \in \mathcal{H}$, we have
\begin{align*}
T_+\varphi\in\ker(T_{\infty} - \zeta_\infty^<)
\ \text{ and } \ 
T_-\varphi\in\ker(T_{-\infty} - \zeta_\infty^>).    
\end{align*}
Thus, $\ker\left(\left( T_{\infty} -\zeta ^{<}_{\infty }\right) T_{+}\right) \cap \ker\left(\left( T_{-\infty} -\zeta ^{>}_{-\infty }\right) T_{-}\right)\neq \{\mathbf{0}\}$.
Furthermore, if (1) and (2) hold, then there exists $\varphi\in\ker\left(\left( T_{\infty} -\zeta ^{<}_{\infty }\right) T_{+}\right) \cap \ker\left(\left( T_{-\infty} -\zeta ^{>}_{-\infty }\right) T_{-}\right)\setminus\{\mathbf{0}\}$ for some $\lambda \in [0,2\pi)$.
Therefore, Corollary \ref{CORO_A} induces eigenvector $\Psi\in\ker(U-e^{i\lambda})$ associated with eigenvalue $e^{i\lambda}$.
\end{proof}

\begin{lemma}
\label{Lemma dimker}
If $e^{i\lambda}\in\sigma_p(U)$, then the followings hold.
\begin{align*}
    &(1) \dim \ker\left(\left( T_{\infty} -\zeta ^{<}_{\infty }\right) T_{+}\right) = \dim \ker\left(\left( T_{-\infty} -\zeta ^{>}_{-\infty }\right) T_{-}\right) = 1.
    \\
    &(2) \ker\left(\left( T_{\infty} -\zeta ^{<}_{\infty }\right) T_{+}\right) = \ker\left(\left( T_{-\infty} -\zeta ^{>}_{-\infty }\right) T_{-}\right).
\end{align*}
\end{lemma}
\begin{proof}
By definition and Theorem \ref{Theorem Ker}, 
\[
1
\leq 
\dim \ker\left(\left( T_{\infty} -\zeta ^{<}_{\infty }\right) T_{+}\right) \leq 2,
\]
\[
1
\leq 
 \dim \ker\left(\left( T_{-\infty} -\zeta ^{>}_{-\infty }\right) T_{-}\right) \leq 2.
\]
Thus, it is sufficient to show that 
\[\dim \ker\left(\left( T_{\infty} -\zeta ^{<}_{\infty }\right) T_{+}\right) = \dim \ker\left(\left( T_{-\infty} -\zeta ^{>}_{-\infty }\right) T_{-}\right) \neq 2.\]
If $\dim \ker\left(\left( T_{\infty} -\zeta ^{<}_{\infty }\right) T_{+}\right) = 2$, then $\left( T_{\infty} -\zeta ^{<}_{\infty }\right) T_{+}$ becomes zero matrix since $T_x$ is $2\times 2$ matrix.
Here, since $T_x$ is a regular matrix, we have
$
\left( T_{\infty} -\zeta ^{<}_{\infty }\right) T_{+} = O 
$
is equivalent to
$
T_{\infty} = \zeta ^{<}_{\infty }
$,
where $O$ is the zero matrix.
This is clearly a contradiction, and $\dim \ker\left(\left( T_{\infty} -\zeta ^{<}_{\infty }\right) T_{+}\right) \neq 2$ holds.
We can show $\dim \ker\left(\left( T_{-\infty} -\zeta ^{>}_{-\infty }\right) T_{-}\right) \neq 2$ by the same argument. Hence(1) is proved.
(2) is immediately shown from (1).
\end{proof}
\begin{corollary}
\label{Coro dimEV}
For $e^{i\lambda}\in\sigma_p(U)$, the following holds.
\[
    \dim\ker(U-e^{i\lambda}) = 1.
\]
\end{corollary}
\begin{proof}
From Theorem \ref{Theorem Ker} and Lemma \ref{Lemma dimker}, we have \[
\dim\ker(U-e^{i\lambda}) = \dim(\ker\left(\left( T_{\infty} -\zeta ^{<}_{\infty }\right) T_{+}\right) \cap \ker\left(\left( T_{-\infty} -\zeta ^{>}_{-\infty }\right) T_{-}\right)) = 1.\]
\end{proof}

\begin{corollary}
\label{Coro_finite}
The following holds.
\[
    \sum_{e^{i\lambda}\in\sigma_p(U)} \dim\ker(U-e^{i\lambda}) < \infty.
\]
\end{corollary}
\begin{proof}
From Theorem \ref{Theorem Ker}, $e^{i\lambda}\in\sigma_p(U)$ has to be a root of the equation
\[
\det\left(\left( T_{\infty} -\zeta ^{<}_{\infty }\right) T_{+}\right) = 0.
\]
Thus, the number of $e^{i\lambda}$ satisfying the above equation is finite.
By combining Corollary \ref{Coro dimEV}, we have the statement.
\end{proof}

\subsection{Strong trapping for quantum walks}
In this subsection, we introduce the mathematical definition of the strong trapping for QWs. The term ``Strong trapping'' was introduced in  \cite{Kollar2015-ar}, and the definition of ``QW is strongly trapped'' is that the QW where a particle starts from the origin shows localization for any initial state.
To prepare for the introduction, we define the time-averaged limit distribution with initial state $\Psi_0\in\mathcal{H}$ as follow:
\[
\overline{\nu}_{\infty } (x)  =\lim _{T\rightarrow \infty }\frac{1}{T}\sum ^{T-1}_{t=0} \| (U^t\Psi_0) (x)\|_{\mathbb{C}^2} ^{2}.
\]
It is well known that the time-averaged limit distribution is written by eigenvalues and eigenvectors of $U$.
For multiplicity $m_\lambda = \dim\ker(U-e^{i\lambda})$ and complete orthonormal basis $\Psi_j^\lambda \in \ker (U-e^{i\lambda}),\ j=1,2,\ldots ,m_\lambda$, the following holds:
\begin{align*}
\overline{\nu}_{\infty } (x) =
\sum _{e^{i\lambda} \in \sigma _{p} (U)}
\sum ^{m_{\lambda }}_{j,k=1}
\overline{\inpr{\Psi^{\lambda }_{k}}{\Psi_0}}
\inpr{\Psi^{\lambda }_{j}}{\Psi_0}
\inpr{\Psi^{\lambda }_{k}(x)}{\Psi^{\lambda }_{j}(x)}
\end{align*}
Moreover, the summation of $\overline{\nu}_{\infty}(x)$ is written by the orthogonal projection onto the direct sum of all eigenspaces $\Pi_p(U) = \sum_{e^{i\lambda}\in\sigma_p(U)} \sum_{j=1}^{m_\lambda}\vert\Psi_j^\lambda\rangle \langle \Psi_j^\lambda\vert$ as follows:
\begin{align*}
\nonumber
    \sum_{x\in\mathbb{Z}} \overline{\nu}_{\infty } (x) &=
    \sum _{e^{i\lambda} \in \sigma _{p} (U)}
\sum ^{m_{\lambda }}_{j,k=1}
\overline{\inpr{\Psi ^{\lambda }_{k}}{\Psi_0}}
\inpr{\Psi^{\lambda }_{j}}{\Psi_0}
\sum_{x\in\mathbb{Z}}
\inpr{\Psi^{\lambda }_{k}(x)}{\Psi^{\lambda }_{j}(x)}
\\
\nonumber
&=
\sum _{e^{i\lambda} \in \sigma _{p} (U)}
\sum ^{m_{\lambda }}_{j,k=1}
\overline{\inpr{\Psi ^{\lambda }_{k}}{\Psi_0}}
\inpr{\Psi^{\lambda }_{j}}{\Psi_0}
\inpr{\Psi^{\lambda }_{k}}{\Psi^{\lambda }_{j}}
\\
\nonumber
&=
\sum _{e^{i\lambda} \in \sigma _{p} (U)}
\sum ^{m_{\lambda }}_{j=1}
\left\lvert\inpr{\Psi^{\lambda }_{j}}{\Psi_0}\right\rvert^2
\\
&=\|\Pi_p(U)\Psi_0\|_{\mathcal{H}}^2.
\end{align*}
Corollary \ref{Coro dimEV} shows that the multiplicity $m_\lambda = 1$ for any $e^{i\lambda}\in\sigma_p(U)$, so the time-averaged limit distributions of our QWs are described as below:
\begin{align*}
\overline{\nu}_{\infty } (x) =
\sum _{e^{i\lambda} \in \sigma _{p} (U)}
\left\lvert\inpr{\Psi^{\lambda }}{\Psi_0}\right\rvert^2\ 
\|\Psi^{\lambda }(x)\|_{\mathbb{C}^2}^2.
\end{align*}
In the following, we assume that the walker starts only from the origin, i.e., the initial state $\Psi_0$ satisfies $\|\Psi_0(0)\|_{\mathbb{C}^2}^2=1$ and $\Psi_0(x)=\mathbf{0}$ if $x\neq 0$.

Richard et al.\cite{Richard2019-fj} and Suzuki \cite{Suzuki2016-qi} showed the weak limit theorem of quantum walks in one dimension with initial state $\Psi_0$ under the short-range assumption:
\[
    C_x = 
    \begin{cases}
    C_{-\infty} +\mathcal{O}(\vert x\vert^{-1-\varepsilon_l}), &\text{ as } x\to -\infty,
    \\
    C_{\infty} +\mathcal{O}(\vert x\vert^{-1-\varepsilon_r}),&\text{ as } x\to \infty.
    \end{cases}
\]
where $\varepsilon_l, \varepsilon_r>0$.
In this weak limit theorem, localization is aggregated into a delta function at the origin with a coefficient $\|\Pi_p(U)\Psi_0\|_{\mathcal{H}}^2$. Thus, it is natural to define the amount of localization by $\|\Pi_p(U)\Psi_0\|^2_\mathcal{H} =
\sum_{x\in\mathbb{Z}}\overline{\nu}_\infty(x)$, and the mathematical definition of strong trapping is introduced as following.
\begin{definition}
A quantum walk is strongly trapped if and only if $\sum_{x\in\mathbb{Z}} \overline{\nu}_\infty(x) \neq 0$ for any $\Psi_0(0)\in\mathbb{C}^2$.
\end{definition}
The following theorem allows us to check whether a quantum walk is strongly trapped or not.
\begin{theorem}
\label{Theorem ST}
A quantum walk is strongly trapped if and only if there exists a pair $e^{i\lambda}, e^{i\lambda'}\in\sigma_p(U)$ such that
$\Psi^{\lambda}(0)$ and $\Psi^{\lambda'}(0)$ are linear independent.
Here,
$\Psi^{\lambda}$ and $\Psi^{\lambda'}$ are eigenvectors associated with $e^{i\lambda}$ and $e^{i\lambda'}$, respectively.
\end{theorem}
\begin{proof}
Firstly, we suppose $\sigma_p(U)\neq\emptyset$ since the quantum walk is obviously not strongly trapped if $\sigma_p(U)=\emptyset$.
By definition, we have
\[
    \overline{\nu}_\infty(x) = \sum_{e^{i\lambda}\in\sigma_p(U)}
    \left\lvert \inpr{\Psi^\lambda(0)}{\Psi_0(0)} \right\rvert ^2\, \|\Psi^\lambda(x)\|^2_{\mathbb{C}^2}.
\]
From Corollary \ref{Coro_finite}, the quantum walk is not strongly trapped, i.e., there exists $\Psi_0(0)\in\mathbb{C}^2$ such that $\sum_{x\in\mathbb{
Z}}\overline{\nu}_\infty(x) = 0$, if and only if there exists $\Psi_0(0)\in\mathbb{C}^2$ which satisfies
\[
\inpr{\Psi^\lambda(0)}{\Psi_0(0)} = 0,\ \text{for all}\ e^{i\lambda}\in\sigma_p(U).
\]
Thus, $\Psi_0(0)$ is orthogonal to all $\Psi^\lambda(0)$. This condition is equivalent to the set of vectors $\{\Psi^\lambda(0)\}_{\Psi^\lambda\in\ker(U-e^{i\lambda})}\subset\mathbb{C}^2$ being linearly dependent. Therefore, the statement is proved.
\end{proof}



\subsection{Two-phase quantum walks with one defect}
In this paper, we consider two-phase QWs with one defect ($x_+ =1,\ x_- = -1$), then   
\begin{align*}
	(\alpha_x, \beta_x, \Delta_x)=
	\begin{cases}
	(\alpha_m, \beta_m, \Delta_m),\quad &x<0,
	\\
	(\alpha_o, \beta_o, \Delta_o),\quad &x=0,
	\\
	(\alpha_p, \beta_p, \Delta_p),\quad &x>0,
	\end{cases}
	\end{align*}
    where $\alpha_j, \beta_j\in\mathbb{C},\ \Delta_j\in[0,2\pi),\ \vert\alpha_j\vert^2+\vert\beta_j\vert^2=1$ and $\alpha_j\neq 0$ for $j\in \{p, o, m\}.$
    Similarly, we write $T_x = T_j,\ \zeta_{x}^\pm=\zeta_{j}^\pm$, where $j=p\ (x>0)$, $=o\ (x=0)$, $=m\ (x<0)$.
    In this case, $T_+$ and $T_-$ equal to $T_o$ and $T_{m}^{-1}$, respectively. 

We now apply Theorem \ref{Theorem Ker} to two-phase QWs with one defect case.
Firstly, we have
\begin{align*}
    \ker\left(\left( T_{\infty} -\zeta ^{<}_{\infty }\right) T_{+}\right) \cap
    \ker &\left(\left( T_{-\infty} -\zeta ^{>}_{-\infty }\right) T_{-}\right)
    \\
    &=
    \ker((T_p - \zeta_p^<)T_{o})
    \cap
    \ker((T_{m} - \zeta_{m}^>)T_{m}^{-1})
    \\
    &=
    \ker((T_p - \zeta_p^<)T_{o})
    \cap
    \ker(T_{m} - \zeta_{m}^>).
\end{align*}
Secondly, $\tilde \Psi$ defined in Corollary \ref{CORO_A} with $\varphi \in \ker((T_p - \zeta_p^<)T_{o})
    \cap
    \ker(T_{m} - \zeta_{m}^>)$ is
\[
\tilde{\Psi } (x)=\begin{cases}
T^{x-1}_{p} T_{o} \varphi,  & x >0,\\
T^{x}_{m} \varphi,  & x\leq 0.
\end{cases} 
\ 
=
\ 
\begin{cases}
\left(\zeta_{p}^{<}\right)^{x-1} T_{o} \varphi,  & x >0,\\
\left(\zeta_{m}^{>}\right)^{x} \varphi,  & x\leq 0.
\end{cases}
\]
Here, we remark that it is sufficient to consider only $\varphi\in\ker(T_m-\zeta^>_m)$ from Lemma \ref{Lemma dimker}.
Finally, eigenvectors $\Psi = J^{-1}\tilde \Psi \in \ker(U-e^{i\lambda})$ is described as follows:

\begin{align*}
\Psi (x) & =\begin{cases}
(\zeta_{p}^{<})^{x-1}
\begin{bmatrix}
\zeta _{p}^{<} & 0\\
0 & 1
\end{bmatrix} T_{o} \varphi,  & x\geq 1,\\[+20pt]
\left(
\begin{bmatrix} 1 & 0 \\ 0 & 0 \end{bmatrix} T_o
+
\begin{bmatrix} 0 & 0 \\ 0 & 1 \end{bmatrix}
\right)\varphi,
 &x=0,\\[+20pt]
(\zeta_{m}^{>})^{x}
\begin{bmatrix}
\zeta _{m}^{>} & 0\\
0 & 1
\end{bmatrix} \varphi,  & x\leq -1.
\end{cases}
\end{align*}

%% file: model.tex
\section{Main Results}
\label{sec:3}    In this section, we analyze eigenvectors and time-averaged limit distributions for concrete five models studied in \cite{Kiumi2021-yg}. Eigenvalues of the time evolution operator have already been given in this study. Note that four of these models are extensions of models in the previous studies\cite{Endo2020-or,Endo2015-db,Endo2014-bu,Endo2015-cy,Wojcik2012-kr}. Eigenvectors of the time evolution operator are derived by Theorem \ref{Theorem Ker}, and we use Theorem \ref{Theorem ST} to clarify whether the model is strongly trapped or not with the initial state $\Psi_0(x) =
    \begin{bmatrix}
    \psi_1 & \psi_2
    \end{bmatrix}^{t}$ with $\vert \psi_1\vert ^2 + \vert \psi_2\vert ^2 = 1$ if $x=0$, and $=
    \begin{bmatrix}
    0 & 0
    \end{bmatrix}^{t}$ if $x\neq 0$.
    Here, $t$ means the transpose operator.
Moreover, an abstract form of time-averaged limit distribution of two-phase QWs with one defect is given in Appendix.

    \begin{model} \label{THEO_MODEL1}
	We assume $(\alpha_p, \beta_p, \Delta_p) = (\alpha_m, \beta_m, \Delta_m) = (\alpha, \beta, \Delta)$ and $\Delta_o = \Delta$. 
	\\
	$\bullet\ ${\bf Eigenvalues :} \\
	$\sigma_p(U)\neq \emptyset$ if and only if $\vert \beta\vert ^2 >\Re(\beta\overline{\beta_o})$ holds, where $\Re$ means the real part of a complex number.
	If this condition holds, the set of eigenvalues is $\sigma_p(U)=\{\pm e^{i\lambda_+},\, \pm e^{i\lambda_-}\}$, where
    \[
    e^{i\lambda_\pm} =\dfrac{A\pm i\sqrt{K}}{\sqrt{A+B}}e^{i\Delta},
    \]
    \[
    A=1-\Re \left( \beta \overline{\beta _{o}}\right),\ B=\vert \beta \vert ^{2} -\Re \left( \beta \overline{\beta _{o}} \right),\ K=\vert \beta \vert ^{2} -\Re ^{2}\left( \beta \overline{\beta _{o}}\right).
    \]
	$\bullet\ ${\bf Eigenvectors :} \\
	The associated eigenvectors of $\pm e^{i\lambda_s}, (s\in\{+,-\})$ are given as below:
\begin{align*}
\Psi _{\pm }^{s} (x)=\pm N\begin{cases}
\dfrac{1}{\alpha _{o}}\begin{bmatrix}
\beta \left( A+si\sqrt{K}\right) +\beta _{o}\left( B-si\sqrt{K}\right)\\
si\dfrac{( A+B)}{\overline{\alpha }}\left(\sqrt{K} -s\Im \left( \beta \overline{\beta _{o}}\right)\right)
\end{bmatrix}\left( \zeta ^{< }\right)^{x} , & x\geq 1,\\[15pt]
\dfrac{1}{\alpha _{o}}\begin{bmatrix}
\beta \left( A+si\sqrt{K}\right) +\beta _{o}\left( B-si\sqrt{K}\right)\\
-\alpha _{o}\left( B-si\sqrt{K}\right)
\end{bmatrix} , & x=0,\\[15pt]
\dfrac{1}{\alpha }\begin{bmatrix}
\beta ( A+B)\\
-\alpha \left( B-si\sqrt{K}\right)
\end{bmatrix}\left( \zeta ^{ >}\right)^{x} , & x\leq -1,
\end{cases}
\end{align*}
where $\Im$ means the imaginary part of a complex number, and
\[
\zeta ^{< } =\pm \dfrac{\overline{\alpha }}{\sqrt{A+B}} ,\qquad \zeta ^{ >} =\pm \dfrac{\sqrt{A+B}}{\alpha },\]
and $N$ is the normalized constant given as below:
\[
    N=\sqrt{\dfrac{\vert  \alpha _{o}\vert  ^{2} B}{2( A+B)^{2}\left(\sqrt{K} -s\Im \left( \beta \overline{\beta _{o}}\right)\right)\sqrt{K}}}.
\]
	\\
	$\bullet\ ${\bf Time-averaged limit distribution :}
\[
\overline{\nu }_{\infty } (x)=2\left(\dfrac{B}{A+B}\right)^{2} \ \begin{cases}
\dfrac{AC_{+}( \psi _{1} ,\psi _{2})}{\vert  \alpha \vert  ^{2} K}\left(\dfrac{\vert  \alpha \vert  ^{2}}{A+B}\right)^{x}, & x\geq 1,\\
1, &x=0,\\
\dfrac{AC_{-}( \psi _{1} ,\psi _{2})}{\vert  \alpha \vert  ^{2} K}\left(\dfrac{\vert  \alpha \vert  ^{2}}{A+B}\right)^{-x}, & x\leq -1,
\end{cases}\]
where
\begin{align*}
&C_{+}(\psi_1,\psi_2) =\vert  \alpha _{o}\vert  ^{2} \vert \beta \vert ^{2} +2\Im ^{2}\left( \beta \overline{\beta _{o}}\right)\vert  \psi _{1}\vert  ^{2} +2\Im \left( \beta \overline{\beta _{o}}\right) \Im \left( \alpha _{o}\overline{\beta } \psi _{1}\overline{\psi _{2}}\right),
\\
&C_{-}(\psi_1,\psi_2) =\vert  \alpha _{o}\vert  ^{2} \vert \beta \vert ^{2} +2\Im ^{2}\left( \beta \overline{\beta _{o}}\right)\vert  \psi _{2}\vert  ^{2} -2\Im \left( \beta \overline{\beta _{o}}\right) \Im \left( \alpha _{o}\overline{\beta } \psi _{1}\overline{\psi _{2}}\right).
\end{align*}
	\\
	$\bullet\ ${\bf Strong trapping : }This model is strongly trapped.
	\begin{proof}
	For instance, $\Psi _{+}^{+} (0)$ and $\Psi _{+}^{-} (0)$ are linear independent. Hence, Theorem \ref{Theorem ST} shows the claim.
	\end{proof}
	
\noindent$\bullet\ ${\bf Example : }
Figure \ref{fig:1} shows the example.
\end{model}

\begin{model}
	\label{THEO_MODEL2}We assume $(\alpha_p, \beta_p, \Delta_p) = (\alpha_m, \beta_m, \Delta_m) = (\alpha, \beta, \Delta)$ and $ \beta_o =  \beta$.
	\\
	$\bullet\ ${\bf Eigenvalues :} \\
	$\sigma_p(U)\neq \emptyset$ if and only if 
	the following Condition 2a or Condition 2b holds:
	\begin{enumerate}
	\rm
	    \item[] Condition 2a : $\gamma_+ = \vert \beta \vert \cos( \Delta -\Delta _{o}) +\vert \alpha \vert \sin( \Delta -\Delta _{o})<\vert \beta\vert $,
	    \item[] Condition 2b : $\gamma_- = \vert \beta \vert \cos( \Delta -\Delta _{o}) -\vert \alpha \vert \sin( \Delta -\Delta _{o})<\vert \beta\vert $.
	\end{enumerate}
	The set of eigenvalues $\sigma_p(U)$ is given as follows:
	\[
	    \sigma_p(U)
	    =
	    \begin{cases}
	        \{\pm e^{i\lambda_+}\},&\, \text{2a holds and 2b does not,} 
	        \\
	        \{\pm e^{i\lambda_-}\},&\,  \text{2b holds and 2a does not,} 
	        \\
	        \{\pm e^{i\lambda_+},\pm e^{i\lambda_-}\},
	        &\,  \text{Both conditions hold,}
	    \end{cases}
	\]
	where
	\[
	 e^{i\lambda _{\pm}} =\dfrac{e^{i\Delta } -\vert \beta \vert ( \vert \beta \vert \pm i\vert \alpha \vert ) e^{i\Delta _{o}}}{\vert e^{i\Delta } -\vert \beta \vert ( \vert \beta \vert \pm i\vert \alpha \vert ) e^{i\Delta _{o}} \vert }.
	\]
	$\bullet\ ${\bf Eigenvectors :} \\
    The associated eigenvectors of $\pm e^{i\lambda _{s}},(s\in\{+,-\})$ are as below:
    \begin{align*}
\Psi _{\pm }^{s} (x)=\pm N\begin{cases}
\dfrac{-si\vert \alpha \vert \vert \beta \vert }{\alpha _{o}\overline{\alpha }\overline{\beta }}\begin{bmatrix}
\overline{\alpha } \vert \beta \vert \left( \vert \beta \vert -( \vert \beta \vert -is\vert \alpha \vert ) e^{i( \Delta -\Delta _{o})}\right)\\
\overline{\beta }\left( 1-2\vert \beta \vert \gamma _{s} +\vert \beta \vert ^{2}\right)
\end{bmatrix}\left( \zeta ^{< }\right)^{x} , & x\geq 1,\\[15pt]
\dfrac{-\vert \beta \vert \left( \vert \beta \vert -( \vert \beta \vert -is\vert \alpha \vert ) e^{i( \Delta -\Delta _{o})}\right)}{\alpha _{o}\overline{\beta }}\begin{bmatrix}
si\vert \alpha \vert \vert \beta \vert \\
\alpha _{o}\overline{\beta }
\end{bmatrix} , & x=0,\\[15pt]
\dfrac{1}{\alpha }\begin{bmatrix}
\beta \left( 1-2\vert \beta \vert \gamma _{s} +\vert \beta \vert ^{2}\right)\\
-\alpha \vert \beta \vert \left( \vert \beta \vert -( \vert \beta \vert -is\vert \alpha \vert ) e^{i( \Delta -\Delta _{o})}\right)
\end{bmatrix}\left( \zeta ^{ >}\right)^{x} , & x\leq -1,
\end{cases}
    \end{align*}
where
\[
\zeta ^{< } =\pm \dfrac{\overline{\alpha }}{\sqrt{1-2\vert \beta \vert \gamma _{s} +\vert \beta \vert ^{2}}} ,\ \ \zeta ^{ >} =\pm \dfrac{\sqrt{1-2\vert \beta \vert \gamma _{s} +\vert \beta \vert ^{2}}}{\alpha },\]
and $N$ is the normalized constant given as below:
\[
   N=\dfrac{\sqrt{( \vert \beta \vert -\gamma _{s})}}{\sqrt{2\vert  \beta \vert  }\left( 1-2\vert \beta \vert \gamma _{s} +\vert \beta \vert ^{2}\right)}.
\]
$\bullet\ ${\bf Time-averaged limit distribution :}
\[
\overline{\nu}_{\infty }(x)=I_{+}\overline{\nu}^{+}_{\infty } (x)+I_{-}\overline{\nu}^{-}_{\infty } (x),
\]
where
\[\ I_{+} =\begin{cases}
1, & \gamma_{+} < \vert \beta \vert ,\\
0, & \gamma_{+} \geq \vert \beta \vert ,
\end{cases}
\qquad
I_{-} =\begin{cases}
1, & \gamma_{-} < \vert \beta \vert ,\\
0, & \gamma_{-} \geq \vert \beta\vert ,
\end{cases}\]
and
\[
\overline{\nu }_{\infty }^{\pm } (x)=C(\psi_1,\psi_2) \begin{cases}
\dfrac{1-\vert \beta \vert \gamma _{\pm }}{\vert  \alpha \vert  ^{2}}\left(\dfrac{\vert  \alpha \vert  ^{2}}{1-2\vert \beta \vert \gamma _{\pm } +\vert \beta \vert ^{2}}\right)^{x} , & x\geq 1,\\
1, & x=0,\\
\dfrac{1-\vert \beta \vert \gamma _{\pm }}{\vert  \alpha \vert  ^{2}}\left(\dfrac{\vert  \alpha \vert  ^{2}}{1-2\vert \beta \vert \gamma _{\pm } +\vert \beta \vert ^{2}}\right)^{-x} , & x\leq -1,
\end{cases}
\]
where
\[
   C( \psi _{1} ,\psi _{2}) =\dfrac{\vert \beta \vert ( \vert \beta \vert -\gamma _{\pm })^{2}\left(\vert  \alpha \vert  \vert \beta \vert \pm 2\Im \left( \alpha _{0}\overline{\beta } \psi _{1}\overline{\psi _{2}}\right)\right)}{\vert  \alpha \vert  \left( 1-2\vert \beta \vert \gamma _{\pm } +\vert \beta \vert ^{2}\right)^{2}}.
\]
$\bullet\ ${\bf Strong trapping :} This model is strongly trapped only in the case both Condition 2a and Condition 2b hold.
In the other cases, the model is {\bf not} strongly trapped.
\begin{proof}
If only one of the conditions holds, then $\Psi_{+}^{s}(0)$ and $\Psi_{-}^{s}(0)$ are obviously linear dependent, and Theorem \ref{Theorem ST} shows the QW is not strongly tapped.
On the other hand, if both conditions hold, then $\Psi_{+}^{+}(0)$ and $\Psi_{+}^{-}(0)$ are linear independent, so this QW is strongly trapped.
\end{proof}
\noindent$\bullet\ ${\bf Example : }
Figure \ref{fig:2} shows the example of not strongly trapped case, and Figure \ref{fig:3} shows strongly trapped case.

\end{model}

\begin{model}
	\label{THEO_MODEL3}
	We assume $(\alpha_o, \beta_o, \Delta_o)=(\alpha_p, \beta_p, \Delta_p)$ and $\arg \beta_p = \arg \beta_m$.
	\\
	$\bullet\ ${\bf Eigenvalues :} \\
	$\sigma_p(U)\neq \emptyset$ if and only if $\cos \left(\Delta_{m}-\Delta_{p}\right)<\left\lvert \beta_{m}\right\rvert \left\lvert \beta_{p}\right\rvert -\left\lvert \alpha_{m}\right\rvert \left\lvert \alpha_{p}\right\rvert $ holds.
	If this condition holds, the set of eigenvalues is $\sigma_p(U) = \{\pm e^{i\lambda}\}$, where
	\begin{align*}
	e^{i \lambda}=
	\dfrac{\left\lvert \beta_{p}\right\rvert  e^{i \Delta_{m}}-\left\lvert \beta_{m}\right\rvert  e^{i \Delta_{p}}}{\vert \vert  \beta_{p}\vert e^{i \Delta_{m}}-\vert  \beta_{m}\vert e^{i \Delta_{p}}\vert }.
	\end{align*}
	$\bullet\ ${\bf Eigenvectors :} \\
	The associated eigenvectors of $\pm e^{i\lambda}$ are given as follows:
    \[
 \Psi_\pm (x)=\pm N\begin{cases}
\dfrac{1}{\alpha _{p}}\begin{bmatrix}
\beta _{m}\left( P+\vert  \beta _{p}\vert  \sqrt{K}\right)\\
-\alpha _{p}\vert  \beta _{m} \vert \left( i\sin( \Delta _{p} -\Delta _{m}) +\sqrt{K}\right)
\end{bmatrix}\left( \zeta _{p}^{< }\right)^{x} , & x\geq 0,\\[15pt]
\dfrac{1}{\alpha _{m}}\begin{bmatrix}
\beta _{m}( M+\vert  \beta _{m} \vert \sqrt{K})\\
-\alpha _{m}\vert  \beta _{m} \vert \left( i\sin( \Delta _{p} -\Delta _{m}) +\sqrt{K}\right)
\end{bmatrix}\left( \zeta _{m}^{ >}\right)^{x} , & x\leq -1,
\end{cases}
\]
where
\[
\zeta _{p}^{< } =\pm \dfrac{\left( P+\vert  \beta _{p}\vert  \sqrt{K}\right)}{\alpha _{p}\sqrt{\vert  \beta _{p}\vert  M-\vert  \beta _{m}\vert  P}} ,\ \zeta _{m}^{ >} =\pm \dfrac{\left( M+\vert  \beta _{m} \vert \sqrt{K}\right)}{\alpha _{m}\sqrt{\vert  \beta _{p}\vert  M-\vert  \beta _{m}\vert  P}},\]
and $N$ is the normalized constant given as below:
\[
    N= \dfrac{\sqrt{\vert  \beta _{m}\vert  \sqrt{K}}}{\sqrt{\vert  \beta _{m}\vert  }(\vert  \beta _{p}\vert  M-\vert  \beta _{m}\vert  P)},
\]
Here,
\[
     P=\vert  \beta _{p}\vert  \cos( \Delta _{p} -\Delta _{m}) -\vert  \beta _{m}\vert ,
    \quad
    M=\vert  \beta _{p}\vert  -\vert  \beta _{m}\vert  \cos( \Delta _{p} -\Delta _{m}),
\]
\[
    K =(\cos( \Delta _{p} -\Delta _{m}) -\vert \beta _{p} \vert \vert \beta _{m} \vert -\vert \alpha _{p} \vert \vert \alpha _{m} \vert )(\cos( \Delta _{p} -\Delta _{m}) -\vert \beta _{p} \vert \vert \beta _{m} \vert +\vert \alpha _{p} \vert \vert \alpha _{m} \vert ).
\]
	$\bullet\ ${\bf Time-averaged limit distribution :}
	\[
	\overline{\nu}_{\infty } (x)=
	C( \psi _{1} ,\psi _{2}) \begin{cases}
\dfrac{P( P+\vert \beta _{p}\vert  \sqrt{K})}{\vert  \alpha _{p}\vert  ^{2}}\left\lvert  \zeta _{p}^{< }\right\rvert  ^{2x}, & x\geq 0,\\
\dfrac{M\left( M+\vert \beta _{m} \vert \sqrt{K}\right)}{\vert  \alpha _{m}\vert  ^{2}}\left\lvert  \zeta _{m}^{ >}\right\rvert  ^{2x}, & x\leq -1,
\end{cases}\]
where
\[
C( \psi _{1} ,\psi _{2}) =\dfrac{4\vert  \beta _{m}\vert  \vert  \beta _{p}\vert  ^{2} K\left(\vert  \alpha _{p}\vert  ^{2}\vert  \beta _{m}\vert  (\vert  \beta _{p}\vert  M-\vert  \beta _{m}\vert  P) -2\left( P+\vert  \beta _{p}\vert  \sqrt{K}\right) c(\psi_1,\psi_2) \right)}{\vert  \alpha _{p}\vert  ^{2}(\vert  \beta _{p}\vert  M-\vert  \beta _{m}\vert  P)^{4}},
\]
and
\[
c( \psi _{1} ,\psi _{2}) =\left( \Re \left( \alpha _{p}\overline{\beta _{m}} \psi _{1}\overline{\psi _{2}}\right) -\vert  \beta _{m}\vert  \vert  \beta _{p}\vert  \vert  \psi _{1}\vert  ^{2}\right)\sqrt{K} -\sin( \Delta _{p} -\Delta _{m}) \Im \left( \alpha _{p}\overline{\beta _{m}} \psi _{1}\overline{\psi _{2}}\right).
\]
	$\bullet\ ${\bf Strong trapping :} This model is {\bf not} strongly trapped.
\begin{proof}
$\Psi_{+}(0)$ and $\Psi_{-}(0)$ are obviously linear dependent, so Theorem \ref{Theorem ST} shows the claim.
\end{proof}

\noindent$\bullet\ ${\bf Example : }
Figure \ref{fig:4} shows the example.

    \end{model}

    \begin{model}
	\label{THEO_MODEL4}
	We assume $(\alpha_o, \beta_o, \Delta_o)=(\alpha_p, \beta_p, \Delta_p)$ and $\Delta_p = \Delta_m =\Delta$.
	\\
	$\bullet\ ${\bf Eigenvalues :} \\
	$\sigma_p(U)\neq \emptyset$ if and only if $PM>0$ holds,
	where
    \[
    P=\vert  \beta _{p}\vert  ^{2} -\Re \left( \beta _{m}\overline{\beta _{p}}\right) ,
    \qquad
    M=\vert  \beta _{m}\vert  ^{2} -\Re \left( \beta _{m}\overline{\beta _{p}}\right).
    \]
    If this condition holds, then the set of eigenvalues is $\sigma_p(U) = \{\pm e^{i\lambda}\}$, where
    \[
     e^{i \lambda}=  
     \dfrac{e^{i \Delta}\left(\sqrt{K}+i \Im\left(\beta_{m} \overline{\beta_{p}}\right)\right)}{\left\lvert \beta_{p}-\beta_{m}\right\rvert }
     ,
    \]
    \[
    K=\left(\Re\left( \beta _{m}\overline{\beta _{p}}\right) +\vert  \alpha _{p}\vert  \vert  \alpha _{m}\vert  -1\right)\left(\Re\left( \beta _{m}\overline{\beta _{p}}\right) -\vert  \alpha _{p}\vert  \vert  \alpha _{m}\vert  -1\right).\]
	$\bullet\ ${\bf Eigenvectors :} \\
	The associated eigenvectors of $\pm e^{i \lambda}$ are given as follows:
    \[
    \ \Psi_\pm (x)=\pm N\begin{cases}
\dfrac{1}{\alpha _{p}}\begin{bmatrix}
-P+\sqrt{K}\\
\alpha _{p}\left(\overline{\beta _{p}} -\overline{\beta _{m}}\right)
\end{bmatrix}\left( \zeta _{p}^{< }\right)^{x}, &x\geq 0,\\[15pt]
\dfrac{1}{\alpha _{m}}\begin{bmatrix}
M+\sqrt{K}\\
\alpha _{m}\left(\overline{\beta _{p}} -\overline{\beta _{m}}\right)
\end{bmatrix}\left( \zeta _{m}^{ >}\right)^{x}, & x\leq -1,
\end{cases}
\]
where
\[
\zeta _{p}^{< } =\pm \dfrac{-P+\sqrt{P+M-\mathfrak{I}^{2}\left( \beta _{m}\overline{\beta _{p}}\right)}}{\alpha _{p}\sqrt{P+M}} ,\ \ \zeta _{m}^{ >} =\pm \dfrac{M+\sqrt{P+M-\mathfrak{I}^{2}\left( \beta _{m}\overline{\beta _{p}}\right)}}{\alpha _{m}\sqrt{P+M}},\]
and $N$ is the normalized constant given as below:
\[
   N=\dfrac{\beta _{m}}{\vert  \beta _{m}\vert  ( P+M)}\sqrt{\dfrac{PM}{\sqrt{K-\mathfrak{I}^{2}\left( \beta _{m}\overline{\beta _{p}}\right))}}}.
\]
$\bullet\ ${\bf Time-averaged limit distribution :}
	\[
\overline{\nu }_{\infty } (x)=C(\psi_1,\psi_2) \begin{cases}
\dfrac{\left(\sqrt{K} -P\right)}{\vert  \alpha _{p}\vert  ^{2}} \ \left\lvert  \zeta _{p}^{< }\right\rvert  ^{2x}, & x\geq 0,\\
\dfrac{\left( M+\sqrt{K}\right)}{\vert  \alpha _{m}\vert  ^{2}}\left\lvert  \zeta _{m}^{ >}\right\rvert  ^{2x}, & x\leq -1,
\end{cases}
\]
where
\[
   C(\psi_1,\psi_2) =\dfrac{4P^{2} M^{2}\left(( P+M) \vert \alpha _{p} \vert ^{2} +2\left( \Re \left( \alpha _{p}\left(\overline{\beta _{p}} -\overline{\beta _{m}}\right) \psi _{1}\overline{\psi _{2}}\right) -P\vert  \psi _{1}\vert  ^{2}\right)\left(\sqrt{K} -P\right)\right)}{( P+M)^{4} \vert \alpha _{p} \vert ^{2}\sqrt{K}} .
\]
	$\bullet\ ${\bf Strong trapping :} This model is {\bf not} strongly trapped.
	\begin{proof}
$\Psi_{+}(0)$ and $\Psi_{-}(0)$ are obviously linear dependent, so Theorem \ref{Theorem ST} shows the claim.
\end{proof}
\noindent$\bullet\ ${\bf Example : }
Figure \ref{fig:5} shows the example.
	\end{model}

	\begin{model}
\label{THEO_MODEL5}
     We assume $\beta_o=0$, $\ \vert \beta_p\vert  = \vert \beta_m\vert =\vert \beta\vert $  and $\Delta_p=\Delta_m =\Delta$.
     \\
    $\bullet\ ${\bf Eigenvalues :} \\
    Let $\gamma=\Delta_{o}+(\arg \beta_p - \arg \beta_{m})/2$.
    Then $\sigma_p(U)\neq\emptyset$ if and only if the following Condition 5a or Condition 5b holds:
    \begin{enumerate}
	\rm
	    \item[] Condition 5a : $\sin( \Delta -\gamma) \in ( -\vert \beta \vert ,1]$,
	    \item[] Condition 5b : $\sin( \Delta -\gamma) \in [ -1,\vert \beta \vert )$.
	\end{enumerate}
	The set of eigenvalues $\sigma_p(U)$ is given as follows:
	\[
	    \sigma_p(U)
	    =
	    \begin{cases}
	        \{\pm e^{i\lambda_+}\},&\quad \text{Condition 5a holds and 5b does not,} 
	        \\
	        \{\pm e^{i\lambda_-}\},&\quad \text{Condition 5b holds and 5a does not,} 
	        \\
	        \{\pm e^{i\lambda_+},\ \pm e^{i\lambda_-}\},
	        &\quad \text{Both conditions hold,}
	    \end{cases}
	\]
	where
	\[
	    e^{i\lambda _\pm} =\dfrac{e^{i\Delta } \pm i\vert \beta \vert e^{i\gamma }}{\vert e^{i\Delta } \pm i\vert \beta \vert e^{i\gamma } \vert }.
	\]
	$\bullet\ ${\bf Eigenvectors :} \\
	The associated eigenvectors of $\pm e^{i\lambda_s},(s\in\{+,-\})$ are given as follows:
	\[
\Psi_\pm^s(x)=\pm N\begin{cases}
\dfrac{1}{\alpha _{o}\overline{\alpha _{p}}}\begin{bmatrix}
\overline{\alpha _{p}} \beta _{m} e^{i( \Delta -\Delta _{0})}\left( 1+si\vert \beta \vert e^{-i( \Delta -\gamma )}\right)\\
si\vert \beta \vert A_s e^{i( \Delta _{o} -\gamma )}
\end{bmatrix}\left( \zeta _{p}^{< }\right)^{x} , & x\geq 1,\\[15pt]
\dfrac{1}{\alpha _{o}}\begin{bmatrix}
\beta _{m} e^{i( \Delta -\Delta _{o})}\left( 1+si\vert \beta \vert e^{-i( \Delta -\gamma )}\right)\\
-\alpha _{o} \vert \beta \vert \left( \vert \beta \vert -sie^{i( \Delta -\gamma )}\right)
\end{bmatrix} , & x=0,\\[15pt]
\dfrac{1}{\alpha _{m}}\begin{bmatrix}
\beta _{m} A_s\\
-\alpha _{m} \vert \beta \vert \left( \vert \beta \vert -sie^{i( \Delta -\gamma )}\right)
\end{bmatrix}\left( \zeta _{m}^{ >}\right)^{x} , & x\leq -1,
\end{cases}
\]
where
\[
A_\pm=1\pm2\vert \beta \vert \sin( \Delta -\gamma ) +\vert \beta \vert ^{2},\ \ \  \zeta _{p}^{< } =\pm \dfrac{\overline{\alpha _{p}}}{\sqrt{A_s}} ,\ \ \zeta _{m}^{ >} =\pm \dfrac{\sqrt{A_s}}{\alpha _{m}}, 
\]
and $N$ is the normalized constant given as below:
\[
    N=\sqrt{\dfrac{\vert \beta \vert +s\sin( \Delta -\gamma )}{2\vert  \beta \vert  A_s^{2}}}.
\]
	$\bullet\ ${\bf Time-averaged limit distribution :} 
	\[
    \overline{\nu}_{\infty } (x)=I_{+}\overline{\nu}^{+}_{\infty } (x)+I_{-}\nu^{-}_{\infty } (x),\]    
where
    \[\ I_{+} =\begin{cases}
1, & \sin( \Delta -\gamma) \in (-\vert \beta \vert ,1],\\
0, &\sin( \Delta -\gamma) \in [ -1,-\vert \beta \vert ],
\end{cases} \ \ \ \ I_{-} =\begin{cases}
1, & \sin( \Delta -\gamma) \in [ -1,\vert \beta \vert ),\\
0, & \sin( \Delta -\gamma) \in [ \vert \beta \vert ,1].
\end{cases}\]
and
\[
\overline{\nu }_{\infty }^{\pm } (x)=C(\psi_1,\psi_2) \begin{cases}
\dfrac{1\pm \vert \beta \vert \sin( \Delta -\gamma )}{\vert  \alpha \vert  ^{2}}\left(\dfrac{\vert  \alpha \vert  ^{2}}{A_\pm}\right)^{x}, & x\geq 1,\\
1 ,& x=0,\\
\dfrac{1\pm \vert \beta \vert \sin( \Delta -\gamma )}{\vert  \alpha \vert  ^{2}}\left(\dfrac{\vert  \alpha \vert  ^{2}}{A_\pm}\right)^{-x}, & x\leq -1,
\end{cases}
\]
where
\[
C(\psi_1,\psi_2) =\dfrac{( \vert \beta \vert \pm \sin( \Delta -\gamma ))^{2}\left(\vert  \beta \vert  ^{2} \mp 2\vert \beta \vert \Im \left( e^{i( \Delta _{o} -\gamma )} \alpha _{o}\overline{\beta _{m}} \psi _{1}\overline{\psi _{2}}\right)\right)}{A_\pm^{2}}.\]
	$\bullet\ ${\bf Strong trapping :} \\ This model is strongly trapped only in the case both Condition 5a and Condition 5b hold.
In the other cases, the QWs is {\bf not} strongly trapped.
\begin{proof}
If only one of the conditions holds, then $\Psi_{+}^{s}(0)$ and $\Psi_{-}^{s}(0)$ are obviously linear dependent, and Theorem \ref{Theorem ST} shows the QW is not strongly tapped.
On the other hand, if both conditions hold, then $\Psi_{+}^{+}(0)$ and $\Psi_{+}^{-}(0)$ are linear independent, so this QW is strongly trapped.
\end{proof}
\noindent$\bullet\ ${\bf Example : }
Figure \ref{fig:6} shows the example of not strongly trapped case, and Figure \ref{fig:7} show strongly trapped case.
    \end{model}

%% file: summary.tex
\section{Summary}
This paper is a continuation of our previous study \cite{Kiumi2021-yg}, which concentrates on the eigenvalues of the time evolution operator for two-phase QWs with one defect. 
In this paper, we focused on the quantitative study of localization and strong trapping property by deriving time-averaged limit distributions on space-inhomogeneous QWs on the integer lattice $\mathbb{Z}$. In Section \ref{sec:2}, we introduced the definitions of our model and gave the method to formulate eigenvectors of time evolution operator. Moreover, we showed some results to characterize eigenvalues not only for two-phase QWs with one defect but also for the more general space-inhomogeneous model. The necessary and sufficient condition for the eigenvectors was shown in Theorem \ref{Theorem Ker}. We also defined the strong trapping with the time-averaged limit distribution, which can be calculated with eigenvectors. In Section \ref{sec:3}, we derived eigenvectors and time-averaged limit distributions for five models whose eigenvalues were derived in the main theorems in the authors' previous study \cite{Kiumi2021-yg}. Models 1 and 2 are one-defect QWs, models 3 and 4 are two-phase QWs, and model 5 are two-phase QWs with one defect. Furthermore, the class of strong trapping was also revealed for these models. 